%% file: paper.tex
\documentclass[aps,prd,nofootinbib,twocolumn,showpacs,superscriptaddress,letterpaper,amsmath,amsfonts,amssymb,preprintnumbers,10pt]{revtex4-2}

\usepackage[bookmarks=false]{hyperref}
\usepackage{graphicx}
\usepackage{bm}
\usepackage{xspace}
\usepackage[separate-uncertainty=true]{siunitx}
\usepackage[italic]{hepnames}
\usepackage{xcolor}
\usepackage{orcidlink}
\usepackage{cleveref}
\usepackage{textcomp}
\usepackage{todonotes}
\usepackage{rotating}
\usepackage{booktabs}
\usepackage{array}
\usepackage{float}
\usepackage{multirow}
\usepackage{placeins}
\usepackage[T1]{fontenc}

\renewcommand{\eqref}[1]{(\ref{#1})}
\crefname{appendix}{Appendix}{Appendices} 
\crefname{equation}{Eq.}{Eqs.}
\crefname{figure}{Fig.}{Figs.} 
\crefname{table}{Table}{Tables}

\newcommand{\dd}{\mathrm{d}}
\newcommand{\acro}[1]{\textsc{\scriptsize #1}\xspace}

\creflabelformat{equation}{#2\textup{#1}#3}

\newcommand{\RDmeas}{\mbox{\ensuremath{
\mathcal{R}(D^+) = 0.418^{+0.075}_{-0.073}~\text{(stat)}^{+0.049}_{-0.056}~\text{(syst)}
}}}

\newcommand{\RDsmeas}{\mbox{\ensuremath{
\mathcal{R}(D^{*+}) = 0.306^{+0.035}_{-0.033}~\text{(stat)}^{+0.016}_{-0.018}~\text{(syst)}
}}}

\newcommand{\lumi}{\mbox{ \ensuremath{365\, \mathrm{fb}^{-1}} }}

\begin{document}

\title{ Test of lepton flavor universality with measurements of $R(D^{+})$ and $R(D^{*+})$ using semileptonic $B$ tagging at the Belle~II experiment}

\input{pub061.tex}   
\begin{abstract}
We report measurements of the ratios of branching fractions \mbox{$\mathcal{R}(D^{(*)+}) = \mathcal{B}(\overline{B}{}^0 \to D^{(*)+} \,\tau^- \, \overline{\nu}_\tau) / \mathcal{B}(\overline{B}{}^0 \to D^{(*)+} \, \ell^- \, \overline{\nu}_\ell)$}, where $\ell$ denotes either an electron or a muon. These ratios test the universality of the charged-current weak interaction. The results are based on a \lumi\ data sample collected with the Belle II detector at the SuperKEKB $e^+e^-$ collider, which operates at a center-of-mass energy corresponding to the $\Upsilon(4S)$ resonance, just above the threshold for $B\overline{B}{}$ production.  Signal candidates are reconstructed by selecting events in which the companion $B$ meson from the $\Upsilon(4S) \to B\overline{B}{}$ decay is identified in semileptonic modes. The $\tau$ lepton is reconstructed via its leptonic decays.  We obtain \RDmeas\ and \RDsmeas, which are consistent with world average values. Accounting for the correlation between them, these values differ from the Standard Model expectation by a collective significance of $1.7$ standard deviations.

\end{abstract}

\maketitle
\twocolumngrid

%%%%%%%%%%%%%%%%%%%%%%%%%%%%%%%%%%%%%%%%%%%%%%%%%%%%%%%%%%%%%%%%%%%%%%%%%%%%%%%%
\section{Introduction}
%%%%%%%%%%%%%%%%%%%%%%%%%%%%%%%%%%%%%%%%%%%%%%%%%%%%%%%%%%%%%%%%%%%%%%%%%%%%%%%%

A fundamental property of the Standard Model (SM) of particle physics is the universality of the electroweak gauge couplings to the three fermion generations. In the lepton sector, this universality results in an accidental symmetry of the lepton flavors that is only broken by the Higgs-Yukawa interaction. One key consequence is that physical processes involving charged leptons feature lepton flavor universality (LFU), an approximate symmetry of lepton flavor among physical observables, only broken by charged lepton mass terms emerging from the nonzero vacuum expectation value of the Higgs field. An observation of lepton flavor universality violation would therefore be a clear signature of physics beyond the SM~\cite{Bernlochner:2021vlv}. 

In this paper, we test LFU using semitauonic \mbox{$b \to c \tau \bar \nu_\tau$} decays by measuring the ratios,\footnote{Charge conjugation is implied throughout this paper.}
\begin{align} \label{eq:RDRDs}
 \mathcal{R}(D^{(*)}) =  \mathcal{R}(D^{(*)+}) &  \equiv \frac{ \mathcal{B}(\overline{B}{}^0 \to D^{(*)+} \tau^- \bar \nu_\tau)  }{  \mathcal{B}( \overline{B}{}^0 \to D^{(*)+} \ell^- \bar \nu_\ell) } \, ,
\end{align}
with $\ell = e, \mu$ and $D^{(*)}$ denoting either $D^{(*)+}$ or $D^{(*)0}$.
The first equality follows from the assumption of isospin symmetry.
Predictions of these ratios are independent of the magnitude of the Cabibbo-Kobayashi-Maskawa (CKM) matrix element $V_{cb}$ and, to some extent, of the parametrization of hadronic matrix elements, reaching a precision of 1--2\%~
\cite{Bigi:2016mdz,Gambino:2019sif,Bordone:2019vic,Martinelli:2021onb,Bernlochner:2022ywh,Ray2024,Aoki2022,PhysRevLett.123.091801,Martinelli:2024epjc}. 
Experimentally, measurements of the ratios in Eq.~\eqref{eq:RDRDs} are preferred to measurements of absolute branching fractions, as efficiency-related systematic uncertainties largely cancel. Furthermore, the simultaneous measurement of both $\mathcal{R}(D)$ and $\mathcal{R}(D^*)$ is useful, as $\overline{B}{} \to D^{*} \tau^- \bar{\nu}_\tau$ decays are an important background in the reconstruction of $\overline{B}{} \to D \tau^- \bar{\nu}_\tau$ decays.

Several experiments previously reported measurements of these or similar ratios~\cite{BaBar:2012obs,BaBar:2013mob,Belle:2015qfa,Sato:2016svk,Belle:2016dyj,Belle:2017ilt,Belle:2019rba,LHCb:2017vlu,LHCb:2022piu,LHCb:2023zxo,LHCb:2023uiv,LHCb:2025fri}.
Belle~II reported \(\mathcal{R}(D^*) = 0.262^{+0.041}_{-0.039}(\mathrm{stat})^{+0.035}_{-0.032}(\mathrm{syst})\)~\cite{PhysRevD.110.072020} and the inclusive ratio $\mathcal{R}(X_{\tau/\ell}) = 0.228 \pm 0.016 \mathrm{(stat)} \pm 0.036 \mathrm{(syst)}$~\cite{Belle-II:2023aih}. A combination of these measurements is reported in Ref.~\cite{HeavyFlavorAveragingGroupHFLAV:2024ctg} and achieves a precision of 8\% for $\mathcal{R}(D)$ and 4\% for $\mathcal{R}(D^*)$, with values of $\mathcal{R}(D) = 0.342 \pm 0.026$ and $\mathcal{R}(D^*) = 0.286 \pm 0.012$. Both exceed the SM expectations of $\mathcal{R}(D) = 0.296 \pm 0.004$ and $\mathcal{R}(D^*) = 0.254 \pm 0.005$~\cite{Bigi:2016mdz,Gambino:2019sif,Bordone:2019vic,Martinelli:2021onb,Bernlochner:2022ywh,Ray2024,Aoki2022,PhysRevLett.123.091801,Martinelli:2024epjc} with a significance of 3.1 standard deviations.

In this analysis, we study $B^0\overline{B}{}^0$ pairs produced in $\Upsilon(4S)$ decays. One $B^0$ or $\overline{B}{}^0$ meson is reconstructed in a semileptonic decay using the hierarchical reconstruction algorithm from Ref.~\cite{Keck:2018lcd}, referred to as $B_{\mathrm{tag}}$. This $B_{\mathrm{tag}}$ is then combined with an oppositely flavored semitauonic or semileptonic decay candidate, which we define as $B_{\mathrm{sig}}$. The selected events are independent of those analyzed in Refs.~\cite{Belle-II:2023aih, PhysRevD.110.072020}, as the latter relied on the reconstruction of hadronically decaying $B_{\mathrm{tag}}$ mesons. The analysis of $B^+B^-$ pairs is deferred to future work, as the reconstruction of the isospin-conjugate $B^-$ signal decays necessitates the efficient identification of $D^{*0} \to D^0 \pi^0$ decays, which suffer from an increased combinatorial background.%, and requires precise efficiency calibrations.

We reconstruct signal candidates from the $D^+ \ell^-$ and $D^{*+} \ell^-$ final states (with $\ell = e, \, \mu$), which can originate from either semitauonic decays ($\overline{B}{}^0 \to D^{(*)+} \tau^- \bar \nu_\tau$ with $\tau^- \to \ell^- \overline{\nu}_\ell \nu_\tau $) or semileptonic decays ($\overline{B}{}^0 \to D^{(*)+} \ell^- \bar \nu_\ell$). These processes differ in the number of neutrinos and are thus distinguishable through kinematic properties. Since both processes produce the same visible final state in the detector, several experimental systematic uncertainties cancel in the measurement of their ratio. 

We distinguish $B_{\mathrm{sig}}$ candidates originating from semitauonic, semileptonic, and background sources using multivariate classifiers trained on the kinematic properties of both the $B_{\mathrm{sig}}$ and $B_{\mathrm{tag}}$ candidates. A binned maximum likelihood fit is then performed to measure the relative contribution from each source, allowing a direct determination of $\mathcal{R}(D^{(*)})$.

The remainder of this paper is structured as follows: Sections~\ref{sec:dataset} and \ref{sec:simulation} provide an overview of the Belle~II detector, the analyzed dataset, and the simulated samples. Sec.~\ref{sec:analysisstrategy} summarizes the tag and signal reconstruction, while Sec.~\ref{sec:signalext} describes the employed multivariate selection. Sec.~\ref{sec:fit} details the fitting procedure, and Sec.~\ref{sec:systematics} discusses the systematic uncertainties affecting the measurement. Sec.~\ref{sec:results} presents our findings and consistency checks, and Sec.~\ref{sec:conclusion} provides our conclusions.

%%%%%%%%%%%%%%%%%%%%%%%%%%%%%%%%%%%%%%%%%%%%%%%%%%%%%%%%%%%%%%%%%%%%%%%%%%%%%%%%
\section{Belle~II detector and dataset}\label{sec:dataset}
%%%%%%%%%%%%%%%%%%%%%%%%%%%%%%%%%%%%%%%%%%%%%%%%%%%%%%%%%%%%%%%%%%%%%%%%%%%%%%%%

The analysis uses Belle~II data collected at SuperKEKB~\cite{Akai:2018mbz} from 2019 to 2022 at a center-of-mass energy of 10.58\,GeV,\footnote{Natural units ($c = \hbar =1$) are used throughout this paper.} corresponding to the $\Upsilon(4S)$ resonance, having an integrated luminosity of 365\,$\text{fb}^{-1}$. The sample contains an estimated $(387 \pm 6) \times 10^6$ $B\overline{B}{}$ events. Additionally, 42.3\,fb$^{-1}$ of off-resonance data at 10.52\,GeV is used to study $q\bar{q}$ ($q = u, d, s, c$) background.  

The Belle~II detector~\cite{Belle-II:2010dht} is an upgraded version of Belle~\cite{Abashian:2000cg} with enhanced particle reconstruction and identification. Its subdetectors are arranged cylindrically around the interaction point (IP), which is enclosed by a \SI{1}{\centi\metre} beryllium beam pipe. The pixel detector (PXD) consists of two layers, with the first fully instrumented and the second partially completed. 
The PXD is surrounded by a four-layer double-sided silicon-strip detector (SVD), and both detectors are used to reconstruct decay vertices with high precision. 
Surrounding these detectors is the central drift chamber (CDC), which provides three-dimensional tracking and specific ionization ($\dd{E}/\dd{x}$) measurements.  

Outside the CDC, the time-of-propagation (TOP) and aerogel ring-imaging Cherenkov (ARICH) detectors provide particle identification in the barrel and forward end cap regions, respectively.

The electromagnetic calorimeter (ECL), consisting of a \SI{3}{\metre} barrel and annular end caps, is located outside the TOP and within a \SI{1.5}{\tesla} superconducting solenoid. The $K^0_L$ and muon detector (KLM), situated outside the solenoid, is composed of iron plates interleaved with active detector elements.

Particle candidates are constructed and identified using the information from various detector systems. 

Charged particle candidates (tracks) are reconstructed by the vertex and tracking systems, and identified based on information from the outer detectors. In particular, muons with sufficiently high momentum will traverse the KLM, while other charged particles are absorbed. In contrast, electrons deposit nearly all of their energy in the ECL. 

Photon candidates consist of ECL clusters that are not consistent with extrapolations of charged tracks. Minimum energy selections are necessary to reject clusters from beam-induced background photons.

%%%%%%%%%%%%%%%%%%%%%%%%%%%%%%%%%%%%%%%%%%%%%%%%%%%%%%%%%%%%%%%%%%%%%%%%%%%%%%%%
\section{Simulation}\label{sec:simulation}
%%%%%%%%%%%%%%%%%%%%%%%%%%%%%%%%%%%%%%%%%%%%%%%%%%%%%%%%%%%%%%%%%%%%%%%%%%%%%%%%

Monte Carlo (MC) samples are used to determine reconstruction efficiencies and acceptance effects as well as to estimate background contamination and to train multivariate classifiers.
The $B$ decays are simulated using the \textsc{EvtGen}\xspace generator~\cite{EvtGen}. The simulation of $e^+ e^- \to q \bar q$ continuum processes is carried out with \acro{KKMC}\xspace~\cite{Jadach:1999vf} and \acro{PYTHIA8}\xspace~\cite{Sjostrand:2014zea}. Electromagnetic final-state radiation is simulated using \acro{PHOTOS}\xspace~\cite{Photos} for all charged final-state particles. Interactions of particles with the detector are simulated using \acro{GEANT4}\xspace~\cite{Agostinelli:2002hh}. The simulated samples contain the equivalent of 2.8\,ab${}^{-1}$ of $B\overline{B}{}$ and continuum processes. The $B\overline{B}{}$ events are simulated with equal fractions of neutral and charged $B$ mesons. An additional sample of $176 \times 10^6$ $\overline{B}{}^0 \to D^{(*)+} \tau^- \bar \nu_\tau$ decays with $\tau^- \to \ell^- \overline{\nu}_\ell \nu_\tau $ is used, corresponding to an effective sample size of $8.5$~ab$^{-1}$.

The signal decays $\overline{B}{}^{0} \to D^{(*)+} \tau^{-} \bar{\nu_\tau}$ and $\overline{B}{}^{0} \to D^{(*)+} \ell^{-} \bar{\nu_\ell}$ are modeled using the form factors from Ref.~\cite{Bernlochner:2022ywh}, with parameter values obtained from a fit to the measurements in Refs.~\cite{Glattauer:2015teq, Waheed:2018djm}. To incorporate this form factor model into the Monte Carlo simulation, the \acro{HAMMER}\xspace software package~\cite{Bernlochner2020} is used to compute and apply event-by-event weights. For the branching fractions isospin-averaged values of Ref.~\cite{HFLAV:2022esi} are used. 

The decays $\overline{B}{}^0 \to D^{**} \tau^-  \overline{\nu}_\tau$ and $\overline{B}{}^0 \to D^{**} \ell \overline{\nu}_\ell$, where $D^{**} = \{D_0^{*+}, D_1^{\prime +}, D_1^+, D_2^{*+} \}$, are modeled using the heavy-quark-symmetry-based form factors proposed in Refs.~\cite{Bernlochner:2016bci,Bernlochner:2017jxt}, with $D^{**}$ masses and widths taken from Ref.~\cite{ParticleDataGroup:2024}. %ParticleDataGroup:2022pth}.
For the $\overline{B}{} \to D^{**} \ell \overline{\nu}_\ell$ branching fractions, we adopt the values from Ref.~\cite{HFLAV:2022esi} to account for missing isospin-conjugated and other established decay modes observed in studies of $B$ decays into fully hadronic final states, following the approach outlined in Ref.~\cite{Bernlochner:2016bci}.

The difference between the inclusive semileptonic branching fraction and the sum of exclusive semileptonic $B$ decays (the so-called ``gap'') is accounted for by using a dedicated sample of $\overline{B}{} \to D^{(*)} (\pi\pi/\eta) \ell \overline{\nu}_\ell$ decays. These are simulated using broad $D^{**}$ contributions that do not form distinct resonance peaks in their invariant mass distributions, resulting in a smooth continuum across phase space. The heavy-quark-symmetry-based form factors of Refs.~\cite{Bernlochner:2016bci,Bernlochner:2017jxt} are used to simulate the decay dynamics. We refer to these as ``nonresonant'' $D^{**}$ decays, in contrast to ``resonant'' $D^{**}$ decays, which proceed via well-defined intermediate states with Breit-Wigner resonance shapes characterized by specific mass and width parameters. Nonresonant $D^{**}$ decays account for approximately $0.7\%$ of all semileptonic and semitauonic events in our reconstructed sample. We also use simulated samples of the isospin-conjugate modes to understand the contamination from $B^+$ decays. 

The simulation is corrected using data-driven weights to account for differences in identification and reconstruction efficiencies. Lepton identification (LID) efficiency and fake rate corrections for electrons are applied as functions of the laboratory-frame momentum, angle relative to the electron beam, and charge of the electron candidate. These corrections are derived from samples of $e^+ e^- \to e^+ e^- (\gamma)$, $e^+ e^- \to e^+ e^- e^+ e^-$, and events with $J/\psi \to e^+ e^-$ decays.  Muon LID corrections are obtained using samples of $e^+ e^- \to \mu^+ \mu^- \gamma$, $e^+ e^- \to e^+ e^- \mu^+ \mu^-$, and events with $J/\psi \to \mu^+ \mu^-$ decays. The rates of misidentifying charged hadrons as leptons are corrected using samples of $K_S^0 \to \pi^+ \pi^-$, $D^{*+} \to D^0 \pi^+$, and $e^+ e^- \to \tau^+ \tau^-$. The efficiency for identifying slow pions from $D^{*+} \to D^0 \pi^+$ decays is corrected using studies of $\overline{B}{}^0 \to D^{*+} \pi^-$. All data and simulated events are reconstructed and analyzed with the open-source basf2 framework~\cite{Kuhr:2018lps}.

%%%%%%%%%%%%%%%%%%%%%%%%%%%%%%%%%%%%%%%%%%%%%%%%%%%%%%%%%%%%%%%%%%%%%%%%%%%%%%%%
\section{Tag and Signal-Side Reconstruction}\label{sec:analysisstrategy}
%%%%%%%%%%%%%%%%%%%%%%%%%%%%%%%%%%%%%%%%%%%%%%%%%%%%%%%%%%%%%%%%%%%%%%%%%%%%%%%%

To select events likely to contain $\Upsilon(4S) \to B^0 \overline{B}{}^0$ decays, we require at least three tracks and three ECL clusters in the event. Tracks must have transverse momenta greater than $0.1$\,GeV and originate within $|d_0| < 0.5$\,cm and $|z_0| < 2.0$\,cm. Here, $|d_0|$ and $|z_0|$ denote the distance of closest approach between the nominal interaction point (IP) and the track in the plane perpendicular to and along the beam axis, respectively. At this stage, all tracks are assigned a pion mass hypothesis. We reconstruct ECL clusters with energy deposits above $0.1$\,GeV that are not associated with any track. Finally, the sum of the selected track and cluster energies must exceed $4$\,GeV.  

We reconstruct $B_{\mathrm{tag}}$ candidates using the \textsc{Full Event Interpretation \acro{(FEI)}}\xspace algorithm~\cite{Keck:2018lcd}. The algorithm constructs $B_{\mathrm{tag}}$ candidates from tracks and clusters, using multivariate classifiers and by using a hierarchical approach. The algorithm is trained to identify semileptonic decays and we use it to reconstruct $B^0 \to D \ell \nu$ and $B^0 \to D^* \ell \nu$ decay candidates, where the $D$ and $D^*$ undergo subsequent hadronic decays. A complete list of decay modes and selection criteria is given in Ref.~\cite{Keck2017_1000078149}. Each $B_{\mathrm{tag}}$ candidate is assigned a confidence score by the algorithm, ranging from zero to one. Candidates with a confidence score above $0.1$ are selected.  
To suppress signal-side semitauonic decays in the $B_{\mathrm{tag}}$ candidates, the lepton momentum in the center-of-mass (c.m.)\ frame must exceed $1$~GeV.  The cosine of the angle between the $B$ meson’s momentum and its visible decay products in the c.m.\ frame is defined as  
\begin{align}\label{eq:costhetby}
\cos\theta_{BY} = \frac{2 E_\mathrm{beam}  E_\mathrm{Y} - m^2_{B} - m^2_\mathrm{Y}}{2 \left|\vec{p}{}_{B}\right| \left|\vec{p}{}_\mathrm{Y}\right|} \, ,
\end{align}
where $E_\mathrm{beam}$ is the beam energy, $m_B$ is the $B$ meson mass, and $|\vec{p}{}_{B}|$ is its momentum, computed from $m_B$ and $E_\mathrm{beam}$. Here $Y = D^{(*)} \ell$ represents the system of visible decay products. For correctly reconstructed semileptonic $B$ decays with a single undetected neutrino, $\cos\theta_{BY}$ lies within $[-1,1]$, but resolution effects and final-state radiation shift it beyond this range. Semitauonic decays with multiple missing neutrinos have on average large negative values of $\cos\theta_{BY}$. All $B_{\mathrm{tag}}$ candidates are required to have $\cos\theta_{BY}$ in the range $[-1.75, 1.1]$, reducing the fraction of semitauonic decays among selected candidates to below $0.4\%$.  

Background from $q\bar{q}$ production is suppressed using Fox-Wolfram moments~\cite{PhysRevLett.41.1581}, which are constructed from a superposition of spherical harmonics using tracks and clusters. The tracks used in the calculation of Fox–Wolfram moments must lie within the CDC acceptance region, have transverse momenta above $0.1$\,GeV, and satisfy $|z_0| < 3.0$\,cm and $|d_0|< 0.5$\,cm. The less stringent $|z_0|$ requirement enhances discrimination against $q\bar{q}$ backgrounds by including tracks with larger longitudinal displacements, which are more characteristic of signal decays and improve event shape information.  We apply a cut on the ratio of the second-order to the zeroth-order Fox–Wolfram moments, with the second-order moment quantifying deviations of the energy flow from isotropy and the zeroth-order moment reflecting the event's spherical geometry. Higher ratios indicate a more collimated structure typical of $q\bar{q}$ events. Therefore, we require this ratio to be less than 0.4 to reduce the $q \bar q$ background.  

On average, approximately 2.02 $B_{\mathrm{tag}}$ candidates per event are reconstructed in the events that passed the selection criteria. From these, we select the candidate with the highest classifier value.  

We reconstruct $B_{\mathrm{sig}}$ candidates in two final states: $D^+ \ell^-$ and $D^{*\, +} \ell^-$ candidates are formed from tracks and clusters not associated with the $B_{\mathrm{tag}}$ candidate. Exactly one charged lepton candidate is required. The signal-side lepton is required to have a charge opposite to the $B_{\mathrm{tag}}$ lepton and is identified using a likelihood-based score that incorporates information from several subdetectors. Electron identification relies on information from the ECL, CDC, TOP, and ARICH, with the most important discriminant being the ratio of reconstructed ECL energy to the estimated track momentum, which is expected to be close to unity for electrons. Electron candidates must have momenta above $0.2$\,GeV, with a loose LID score selection. We correct the electron energy for bremsstrahlung losses by adding back ECL clusters that are near the tracks, following the methodology of Ref.~\cite{Belle-II:2023aih}. Muons are identified by extrapolating tracks to the KLM, where the likelihood is primarily constructed from the longitudinal penetration depth and transverse scattering of the extrapolated track. Muon candidates must have momenta above $0.4$\,GeV, with a stringent LID score requirement. The efficiency for correctly identifying electrons is 98.7\% (99.7\%) in semitauonic (semileptonic) events, with a misidentification rate such that only 1\% of pions or kaons pass this requirement. The efficiency for correctly identifying muons is 79.3\% (86.3\%) in semitauonic (semileptonic) events, with only 5\% of pions or kaons satisfying the selection.

Neutral pion ($\pi^0$) candidates are reconstructed from pairs of ECL clusters not associated with any tracks with an invariant mass between $120$ and $145$\,MeV. To suppress background, clusters must have energies above $0.08$, $0.03$, or $0.06$\,GeV in the forward, barrel, and backward regions, respectively. Each cluster must consist of multiple crystals, lie within the CDC angular acceptance, and have a measured time within $200$\,ns of the expected event time. A multivariate classifier is constructed from electromagnetic cluster shape quantities and combined with the photon’s distance to the nearest track to distinguish real photons from clusters originating from hadronic showers. More details can be found in Ref.~\cite{PhysRevD.110.072020}.

Neutral kaon ($K_S^0$) candidates are reconstructed from pairs of charged particles, each of which is assigned a pion mass hypothesis, whose combined invariant mass lies between $450$ and $550$\,MeV and which can be fit to a common vertex. The flight distance must be positive, the significance of the displacement between the point of closest approach and the IP must exceed $0.5 $%\, \mathrm{cm}$
, and the cosine of the angle between the momentum and vertex position vector must be greater than $0.8$. 
The displacement significance, defined as the separation divided by its uncertainty, distinguishes true $K_S^0$ candidates from random combinations of tracks.

The decays of $D$ mesons are reconstructed in modes with large branching fractions and high purities. For the $D^{+} \ell^-$ final state, we include $D^+ \to K^- \pi^+ \pi^+$, $K^0_S \pi^+ \pi^0$, $K^0_S \pi^+ \pi^+ \pi^-$, $K^0_S \pi^+$, $K^- K^+ \pi^+$, and $K^0_S K^+$. For the $D^{*+} \ell^-$ final state, where only the $D^{*+} \to D^0 \pi^+$ decay is used, we include $D^0 \to K^- \pi^+ \pi^0$, $K^- \pi^+ \pi^+ \pi^-$, $K^0_S K^+ K^-$, $K^+ K^-$, $K^- \pi^+$, $K^0_S \pi^+ \pi^-$, and $\pi^- \pi^+$. All $D$ candidates must have a reconstructed invariant mass within $2.5 \sigma$ of the nominal $D$ mass \cite{ParticleDataGroup:2024}, where $\sigma$ refers to the resolution of the mass peak.

For the reconstruction of $D^{*+}$ candidates, each $D^0$ candidate is combined with a single charged track, assumed to be a pion. Candidates are required to have a mass difference $\Delta m = m(D^{*+}) - m(D^0)$ between 130 and 160\,MeV, corresponding to 2.8 times the $\Delta m$ resolution. An additional vertex fit is performed on each $D^{*+}$ candidate to update its momentum. Using MC simulation, we estimate that for true $D^{*+}$ candidates, the slow pion is correctly identified in 71\% of cases.

The construction of $B_{\mathrm{sig}}$ candidates is achieved by combining $D^{(*)+}$ and $\ell^-$ candidates. The requirement $-15 < \cos\theta_{BY} < 1.1$ selects semitauonic and semileptonic final states, with approximately 5\% of semitauonic signal events falling outside this range on the negative end.

Finally, $B_{\mathrm{tag}}$ and  $B_{\mathrm{sig}}$ candidates are combined to form $\Upsilon(4S)$ candidates, requiring no additional charged tracks in the event. We demand that $B_{\mathrm{tag}}$ and  $B_{\mathrm{sig}}$ are of opposite flavor and reject events in which one $B$ mixed, as such events possess lower purity. On average, fewer than $1.08$ $\Upsilon(4S)$ candidates remain per event after applying the selection criteria, and we select a single candidate per event based on the highest signal-side $D^{(*)+}$ vertex fit $p$-value.  The unassigned energy in the calorimeter of the selected $\Upsilon(4S)$ candidate, $E_\mathrm{extra}$, is calculated by summing clusters not associated with any particles used in the reconstruction. Two multivariate algorithms, described in Ref.~\cite{Cheema:2024iek} and based on the \textsc{FastBDT}\xspace classifier \cite{FastBDT}, remove contributions from beam background and 
hadronic split-offs, by utilizing features based on timing, energy spread, scintillation pulse shape, and cluster localization in the ECL.
Clusters added as bremsstrahlung corrections to tracks are excluded. From correctly reconstructed events we expect values of $E_\mathrm{extra}$ near zero, while background events typically exhibit higher values. We only retain events with $E_\mathrm{extra} < 1.2$\,GeV. 

Appendix~\ref{app:sel} provides more details on the differences of the signal and tag-side selection and reconstruction.

%%%%%%%%%%%%%%%%%%%%%%%%%%%%%%%%%%%%%%%%%%%%%%%%%%%%%%%%%%%%%%%%%%%%%%%%%%%%%%%%
\section{Multivariate Classification}\label{sec:signalext}
%%%%%%%%%%%%%%%%%%%%%%%%%%%%%%%%%%%%%%%%%%%%%%%%%%%%%%%%%%%%%%%%%%%%%%%%%%%%%%%%

Signal extraction is performed using a multiclass classification algorithm that differentiates semitauonic signal, semileptonic signal, and background events. The model employs gradient-boosted decision trees (BDTs), where each tree corrects the errors of the previous ones to improve classification. Further details can be found in Ref.~\cite{scikit-learn}.

\begin{figure}[h!] 
	
     \includegraphics[width=0.27\textwidth,trim=3.2cm 4.5cm 3.5cm 3.5cm, clip]{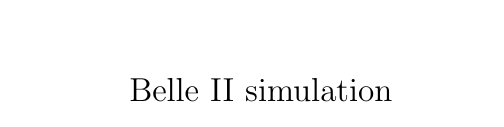} \\
        \includegraphics[width=0.22\textwidth,trim=0cm 0cm 0cm 1cm, clip]{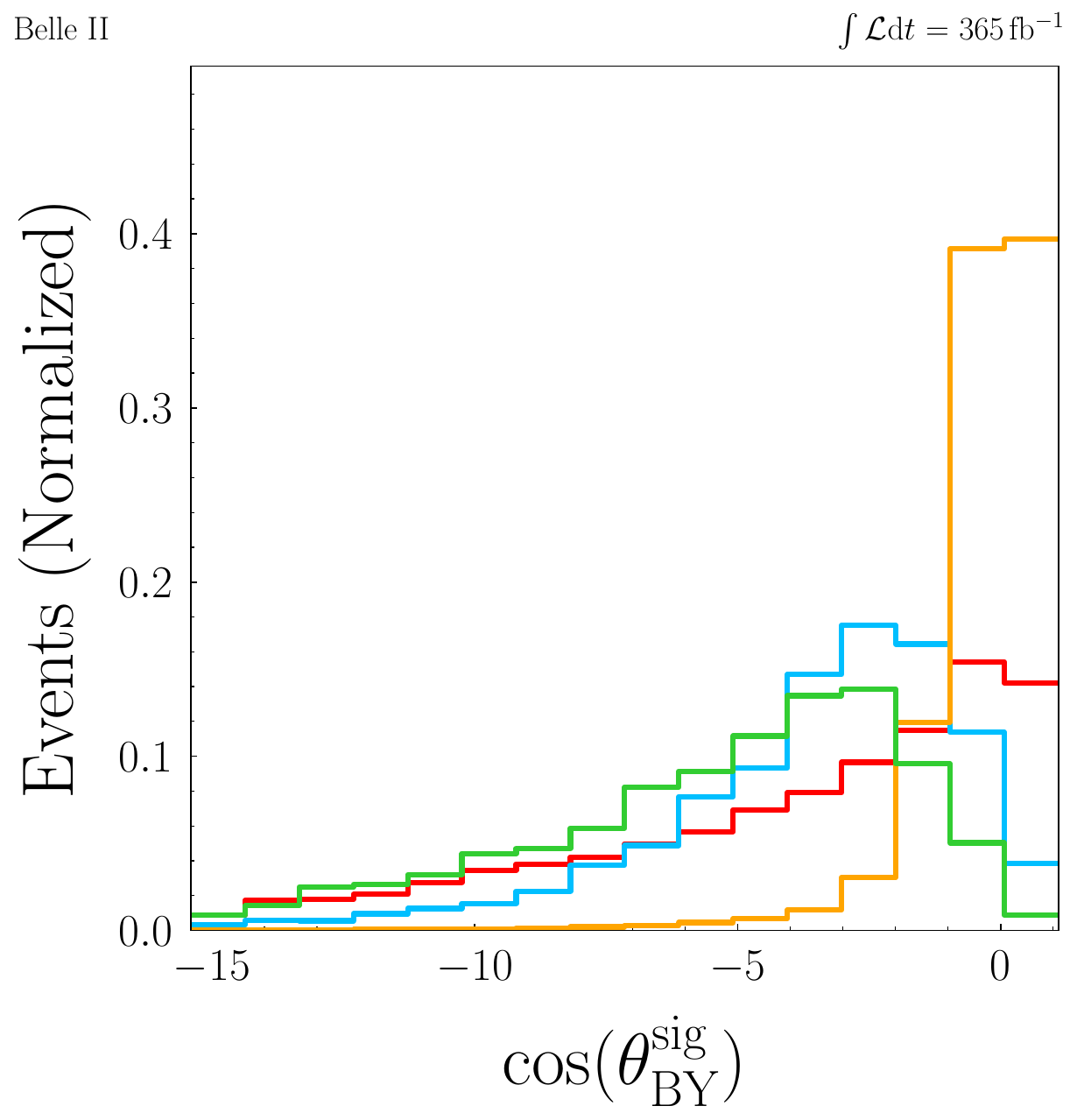}
         \put(-92,78){\includegraphics[width=0.31\linewidth,clip]{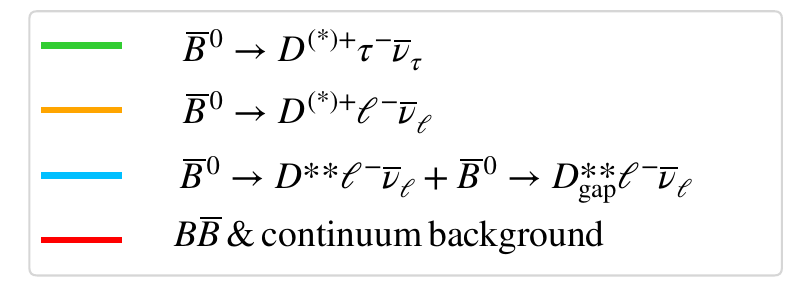}} \quad
        \includegraphics[width=0.22\textwidth,trim=0cm 0cm 0cm 1cm, clip]{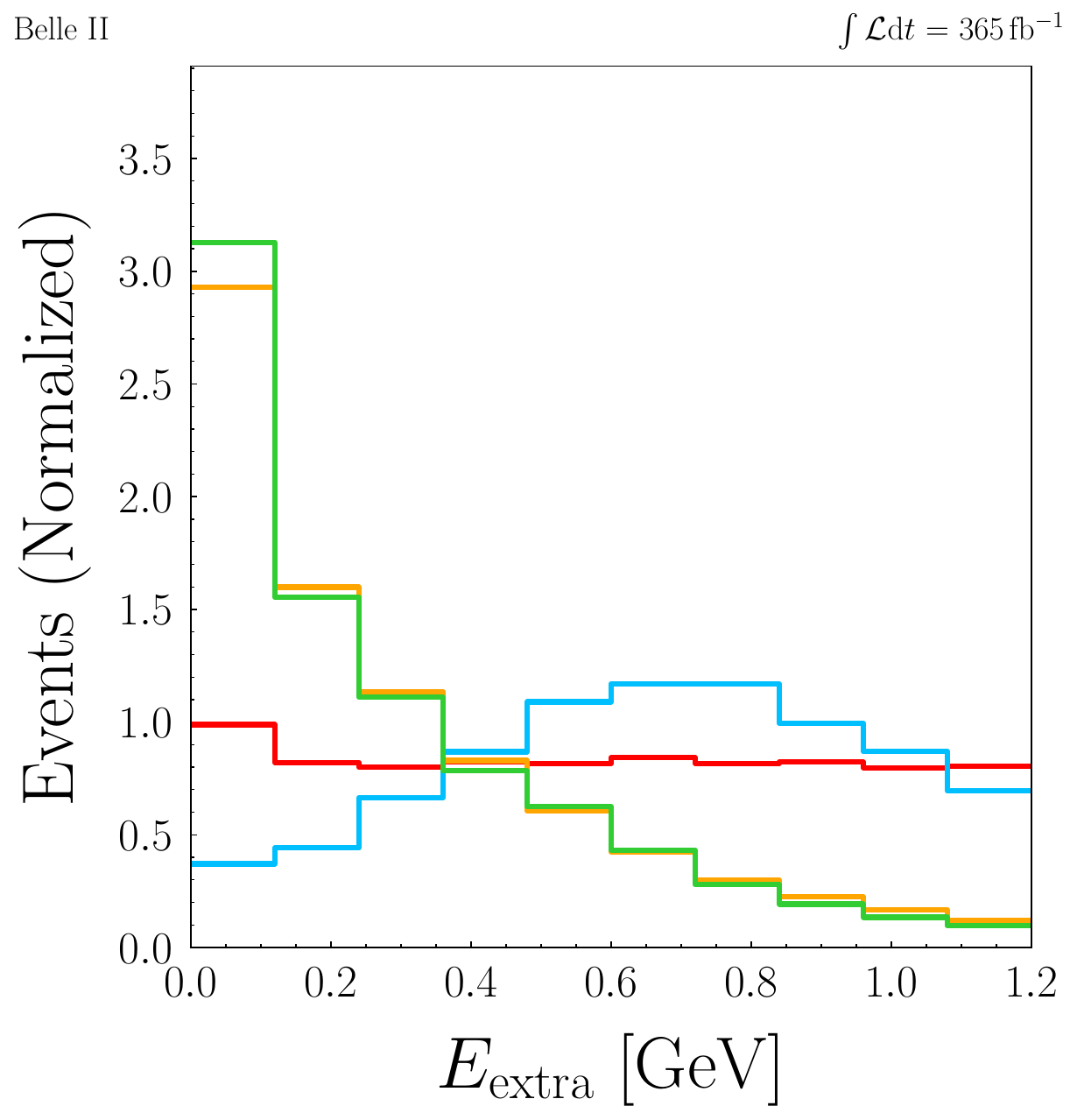}\\
        \vspace{1ex}
        \includegraphics[width=0.22\textwidth,trim=0cm 0cm 0cm 1cm, clip]{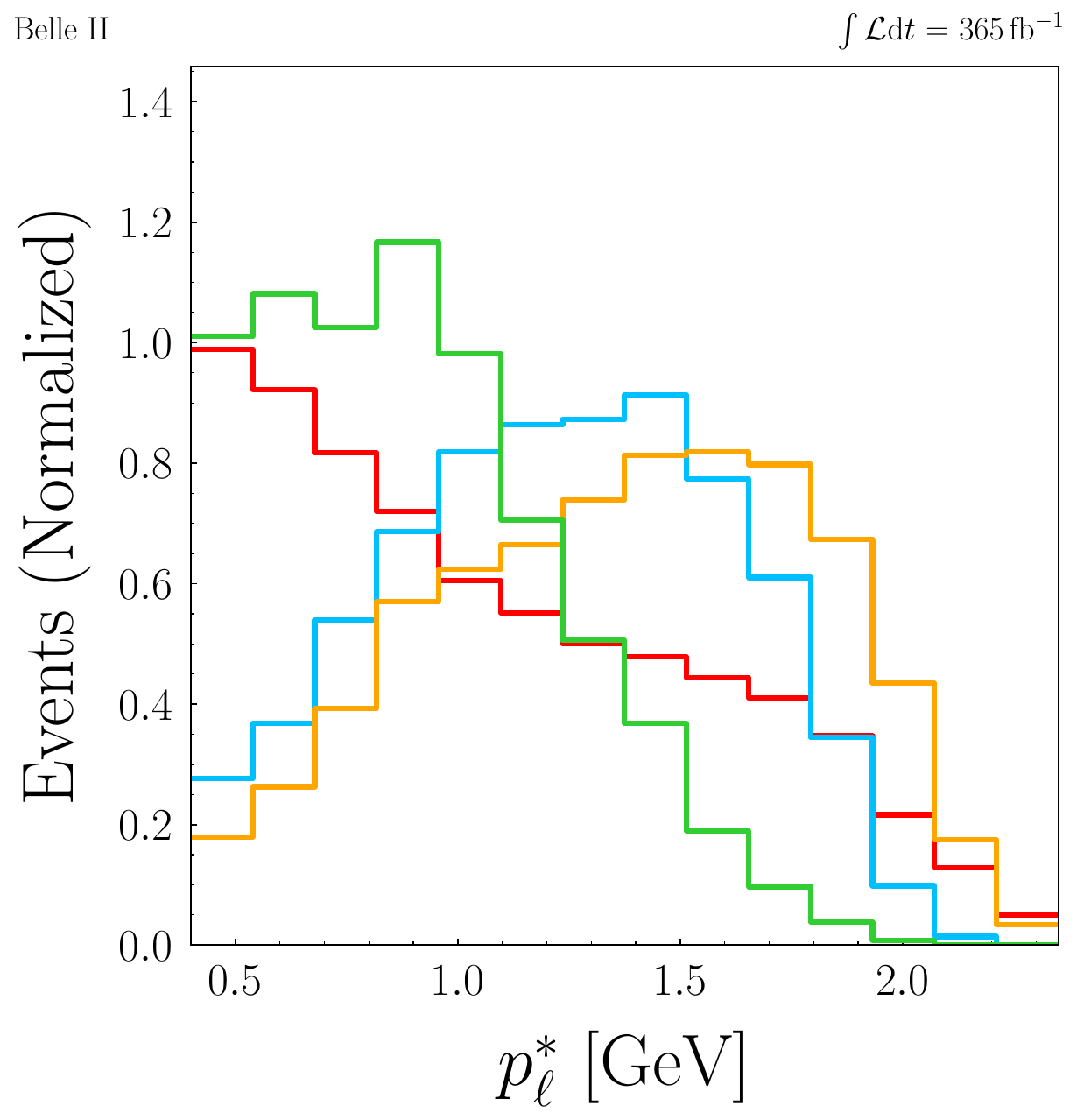} \quad
        \includegraphics[width=0.22\textwidth,trim=0cm 0cm 0cm 1cm, clip]{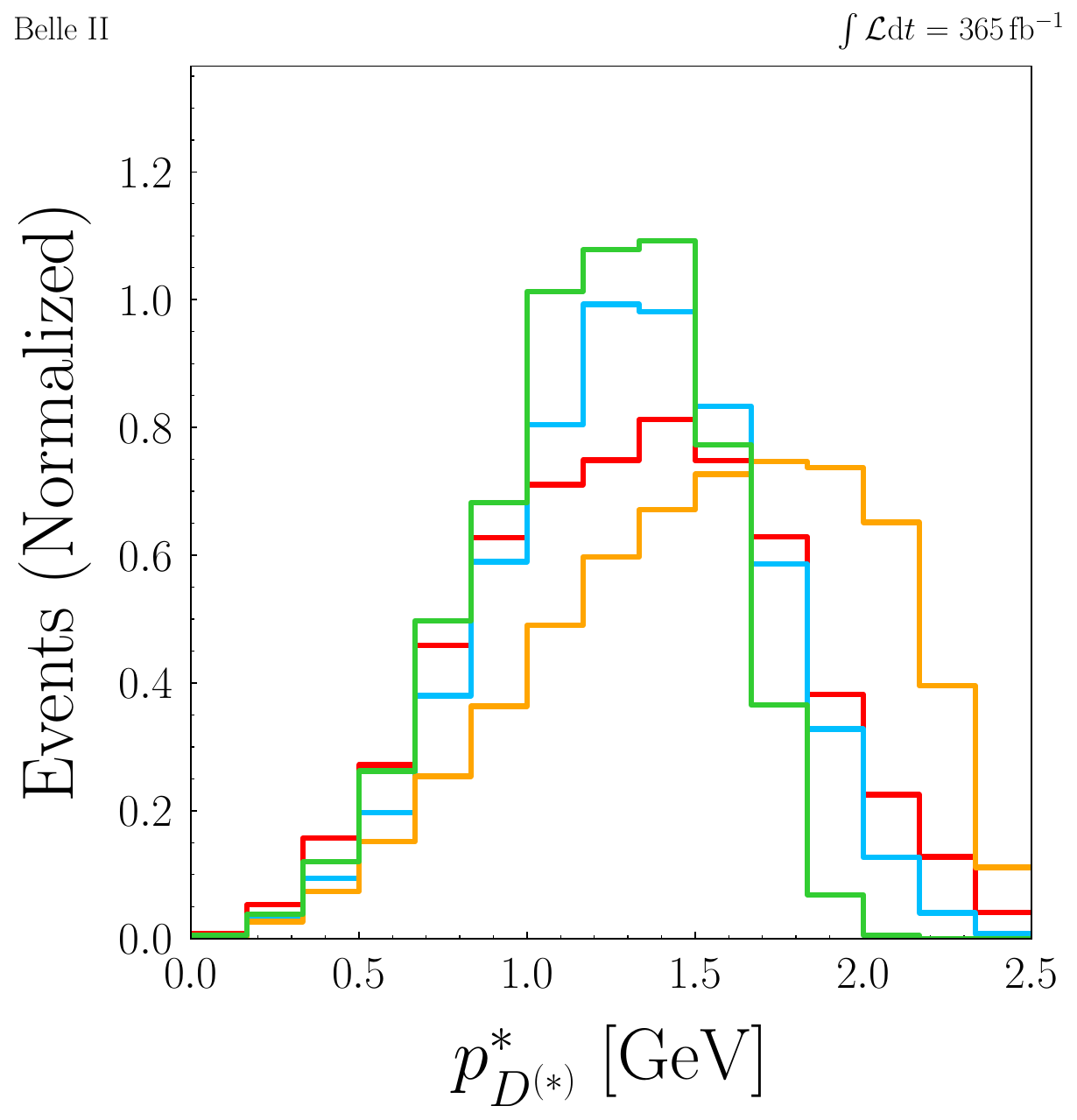} \quad \\   
        \vspace{1ex}        
        \centering
           \includegraphics[width=0.22\textwidth,trim=0cm 0cm 0cm 1cm, clip]{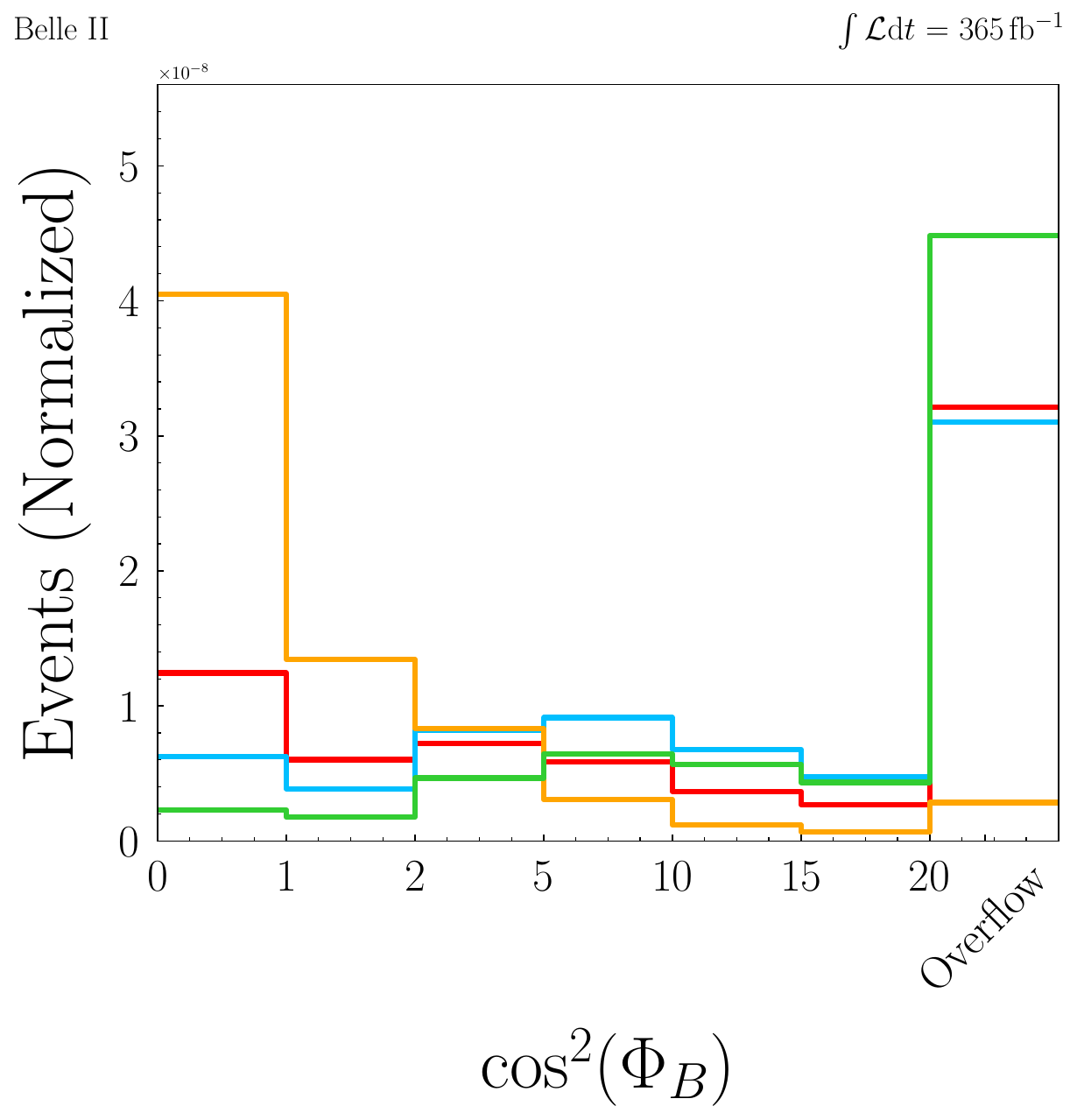} 
	\caption{
	The five input variables for the multiclassification BDT are shown for electrons and muons, and the $D^+\ell$ and $D^{*+} \ell$ categories combined. 
}
	\label{fig:BDT_shapes}
\end{figure}

The BDT is trained on five input variables. The most discriminating variable is $\cos\theta_{BY}$ of the $B_{\mathrm{sig}}$ candidate, followed by the unassigned energy in the calorimeter, $E_{\mathrm{extra}}$. The third most important variable, $\cos^2 \Phi_B$, is defined as
\begin{align}
 \cos^2 \Phi_B = \frac{\cos^2\theta_{BY}^\mathrm{sig} + \cos^2\theta_{BY}^\mathrm{tag} + 2 \cos\theta_{BY}^\mathrm{sig}\cos\theta_{BY}^\mathrm{tag} \cos\gamma}
    {\sin^2 \gamma} \, .
\end{align}  
It combines $\cos\theta_{BY}$ from both the $B_{\mathrm{sig}}$ and $B_{\mathrm{tag}}$ candidates with the angle $\gamma$ between their $Y$ momenta. For correctly reconstructed semileptonic $B_{\mathrm{sig}}$ and $B_{\mathrm{tag}}$ candidates, each with a single missing neutrino, $\cos^2 \Phi_B$ is expected to take values between zero and one. Events with multiple missing neutrinos, such as semitauonic decays or misreconstructed events, tend to have larger values. The fourth and fifth most important input variables are the center-of-mass momenta of the $D$ ($p_D^*$) and lepton ($p_\ell^*$) candidates, respectively. These variables help distinguish semitauonic, semileptonic, and background events based on the different phase space available in each case. Fig.~\ref{fig:BDT_shapes} shows the five input variables for $D$ and $D^*$ candidates with electrons and muons combined. A good separation of all three event types can be obtained between semileptonic and semitauonic signal decays in $\cos\theta_{BY}$, $\cos^2 \Phi_B$, $p_\ell^*$, and $p_D^*$, whereas $E_{\mathrm{extra}}$ is very powerful at separating semileptonic and semitauonic signal from other semileptonic processes or backgrounds. The three resulting classification scores are denoted as $z_\tau$, $z_{\ell}$, and $z_\mathrm{bkg}$ for semitauonic, semileptonic, and background events, respectively. 

The performance of the classification was tested using independent MC events, and no signs of overtraining were detected. The MC events used for training and testing were not used further in the analysis.

%%%%%%%%%%%%%%%%%%%%%%%%%%%%%%%%%%%%%%%%%%%%%%%%%%%%%%%%%%%%%%%%%%%%%%%%%%%%%%%%
\section{Fitting Procedure}\label{sec:fit}
%%%%%%%%%%%%%%%%%%%%%%%%%%%%%%%%%%%%%%%%%%%%%%%%%%%%%%%%%%%%%%%%%%%%%%%%%%%%%%%%

We extract the signal using a binned two-dimensional log-likelihood fit to the variables $z_{\tau}$ and $z_\mathrm{diff} = z_{\ell} - z_\mathrm{bkg}$. We consider four categories of events: $D^+e^-$, $D^{*+}e^-$, $D^+\mu^-$, and  $D^{*+}\mu^-$ candidates. The likelihood is implemented using the \textsc{pyhf}\xspace package~\cite{pyhf,Cranmer:1456844}. The total likelihood function has the form
\begin{equation} \label{eq:likelihood}
 \mathcal{L} = \prod_c \, \mathcal{L}_c \, \times \prod_l \, \mathcal{G}_l \, ,
\end{equation}
with the individual category likelihoods $\mathcal{L}_c$ and nuisance-parameter (NP) constraints $\mathcal{G}_l$. The product in Eq.~\eqref{eq:likelihood} runs over all categories $c$ and independent uncertainty sources $l$, respectively. The role of the NP constraints is detailed in Sec.~\ref{sec:systematics}. Each category likelihood $\mathcal{L}_c$  is defined as the product of individual Poisson distributions $\mathcal{P}$,
\begin{align}
   \mathcal{L}_c = \prod_i^{\rm bins} \, \mathcal{P}\left( n_i ; \nu_i \right)  \, 
\end{align}
with $n_i$ denoting the number of observed data events and $\nu_i$ the total number of expected events in a given bin $i$. The number of expected events in a given bin, $\nu_i$, is estimated using simulated events. It is given by
\begin{equation}\label{eq:nui}
 \nu_i = \sum_k^{\rm processes} \, \eta_k \, f_{ik}   \, ,
\end{equation}
where $\eta_k$ is the total number of events from a given process $k$ with a fraction $f_{ik}$ of such events being reconstructed in the bin $i$. The values of $\cal{R}(D^+)$ and $\cal{R}(D^{*+})$ and the sum of semileptonic signal decays $\eta_{D^{(*)}\ell}$ from electrons and muons are used to determine the number of semitauonic decays $\eta_{D^{(*)}\tau}$ via
\begin{align} \label{eq:RD_RDs_fit}
 \eta_{D^{(*)}\tau} = \frac 1 2 \cal{R}(D^{(*)+}) \times \eta_{D^{(*)}\ell} \times \left( \frac{ \epsilon_{D^{(*)}\tau} }{  \epsilon_{D^{(*)}\ell}} \right) \, ,
\end{align}
where $\epsilon_{D^{(*)}\tau}$ and $\epsilon_{D^{(*)}\ell}$ denote the efficiencies of semitauonic and semileptonic signal decays from electrons and muons.

\begin{figure*}[!htbp] 
	\centering
        \includegraphics[width=0.75\textwidth, trim=0.5cm 0.5cm 0.5cm 0.5cm, clip]{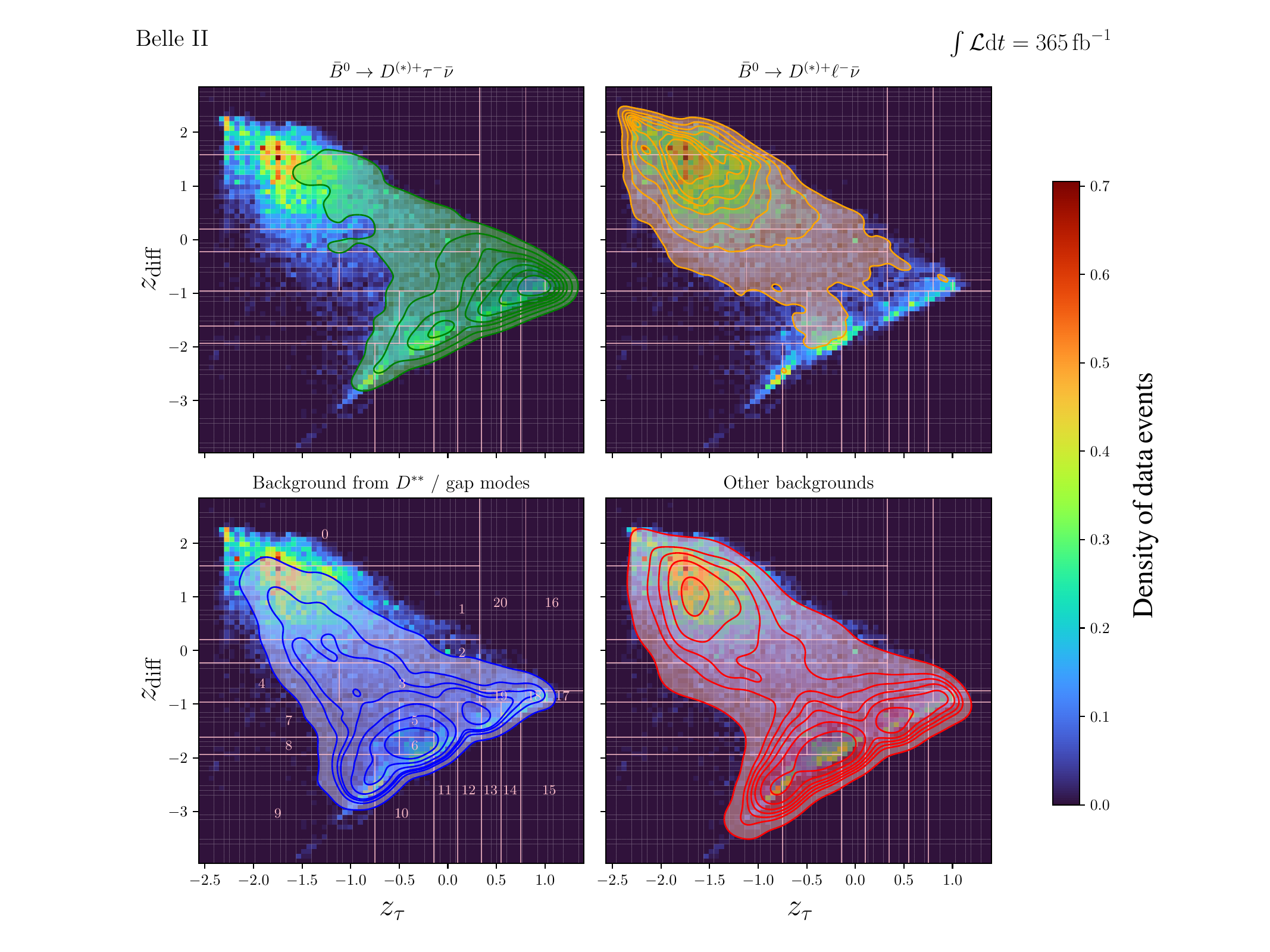}
	\caption{
        The densities of the four fit categories in the $z_\mathrm{diff} = z_{\ell} - z_\mathrm{bkg}$ vs. $z_\tau$ plane and the chosen binning are shown. The four components, shown from top left to bottom right, are semitauonic, semileptonic, $D^{**}$ and gap-mode backgrounds, and other backgrounds. The binned histogram shows the density of data events. 
}
	\label{fig:binning}
\end{figure*}

The factor of $\frac{1}{2}$ in Eq.~\eqref{eq:nui} accounts for the definition of the semileptonic signal, which includes both electron and muon final states. We implement a non-uniform binning scheme in both dimensions, increasing granularity in regions where the classifier is most sensitive to semitauonic events and $\overline{B}{} \to D^{**} \ell \bar{\nu}_\ell$ as well as other semileptonic gap backgrounds. In contrast, the region dominated by semileptonic $\overline{B}{} \to D^{(*)} \ell \bar{\nu}_\ell$ events is binned more coarsely, as these events are readily identifiable. Fig.~\ref{fig:binning} shows the binning structure for semitauonic and semileptonic events, for the background category from $\overline{B}{} \to D^{**} \ell \bar{\nu}_\ell$ and gap modes, and for other backgrounds.  

The likelihood (Eq.\eqref{eq:likelihood}) is numerically maximized to fit the values of the different components, $\eta_k$, using the observed events. This maximization is performed with the \textsc{iMinuit}\xspace package\cite{iminuit}. The ten parameters of interest we determine are
\begin{itemize}
 \item $\mathcal{R}(D^+)$ and $\mathcal{R}(D^{*+})$ (2 parameters, shared between $e$ and $\mu$ channels);
 \item Normalizations $\eta_{D^{(*)}\ell}$ of \mbox{$\overline{B}{}^{0} \to D^+ \ell^- \bar \nu_\ell$} and \mbox{$\overline{B}{}^{0} \to D^{*+} \ell^- \bar \nu_\ell$} events (2 parameters, shared between $e$ and $\mu$ channels);
 \item Background from $B \to D^{**} \ell \bar \nu_\ell$ and semileptonic gap events (2 parameters; shared between $e$ and $\mu$ channels);
 \item Number of other $B \overline{B}{}$ or continuum background events (4 parameters; one for each fit category).
\end{itemize}

We also test LFU between electrons and muons for the semileptonic signal modes. For this we modify the fit setup by introducing four normalization parameters for each of the four decays of interest and modify Eq.~\eqref{eq:RD_RDs_fit} accordingly.

To validate the fit procedure, we generated ensembles of pseudoexperiments for different input $\mathcal{R}(D^+)$ and $\mathcal{R}(D^{*+})$ values. Fits to these ensembles show no biases in central values and no under- or overcoverage of determined confidence intervals. Assuming SM values for $\mathcal{R}(D^+)$ and $\mathcal{R}(D^{*+})$, we expect significances of 3.9 and 7.0 standard deviations, respectively. 

%%%%%%%%%%%%%%%%%%%%%%%%%%%%%%%%%%%%%%%%%%%%%%%%%%%%%%%%%%%%%%%%%%%%%%%%%%%%%%%%
\section{Systematic Uncertainties}\label{sec:systematics}
%%%%%%%%%%%%%%%%%%%%%%%%%%%%%%%%%%%%%%%%%%%%%%%%%%%%%%%%%%%%%%%%%%%%%%%%%%%%%%%%

Several systematic uncertainties affect the measured ratios of $\mathcal{R}(D^+)$ and $\mathcal{R}(D^{*+})$, as summarized in Table~\ref{tab:systematics}, with a corresponding table for the LFU tests provided in the Appendix (Table~\ref{tab:systematics_lfu}).

The effect of systematic uncertainties is directly incorporated into the likelihood function. We distinguish between additive and multiplicative uncertainties: Additive uncertainties affect the signal and background template shapes, whereas multiplicative uncertainties affect efficiencies or branching fractions of semitauonic and semileptonic signal events. A vector of NPs, $\boldsymbol{\theta}$,  is introduced and each NP is constrained in the likelihood Eq.~\eqref{eq:likelihood} using a standard normal distribution $\mathcal{G}_l = \mathcal{G}(\theta_l)$ with $l$ denoting an independent uncertainty source or if applicable Poisson constraints.

A brief summary of each significant uncertainty source follows, ordered by their importance: 

The dominant uncertainty arises due to the finite size of the simulated data samples and is estimated using the ``Barlow-Beeston Lite'' method~\cite{Cranmer:1456844}, introducing Poisson constraints to correctly treat sparsely populated bins.

The next largest uncertainty stems from the limited knowledge on the composition and modeling of the semileptonic gap processes. To account for the former, we assign a 100\% uncertainty to their branching fractions. The latter is addressed by varying the heavy-quark-symmetry-based form factors independently for each assumed gap process, using the uncertainties and correlations provided in Refs.~\cite{Bernlochner:2016bci,Bernlochner:2017jxt} for the broad states. Due to the larger contamination, their impact is more pronounced in the $D$ channel than in the $D^*$ channel, translating into a larger systematic uncertainty for $\mathcal{R}(D^+)$. The resulting systematic uncertainty is inherently asymmetric, as the gap event yield is constrained to be non-negative in the fit.

In contrast, the branching fractions and modeling of resonant $\overline{B}{} \to D^{**} \ell \bar{\nu}_\ell$ decays are better constrained experimentally~\cite{ParticleDataGroup:2022pth} and we vary the form factors of the broad and narrow states using the uncertainties and correlations provided in Refs.~\cite{Bernlochner:2016bci,Bernlochner:2017jxt}. 

Lepton identification impacts the precision of $\mathcal{R}(D^+)$ and $\mathcal{R}(D^{*+})$ due to the limited size and systematic uncertainties of the calibration samples used to correct discrepancies between data and Monte Carlo simulations. These effects influence both correctly identified leptons and misidentified leptons (``fakes''). The contribution of misidentified background events is larger in the signal-enriched regions of the $D (e/\mu)$ channels compared to the $D^{*} (e/\mu)$ channels, resulting in a higher uncertainty in $\mathcal{R}(D^+)$. The additive lepton identification uncertainties can become asymmetric due to correlations with background yields.

We assign a track reconstruction efficiency uncertainty of 0.3\% per track for kaon, pion, and lepton tracks, reflecting the imperfect knowledge of the track-reconstruction efficiency. This uncertainty is estimated using a control sample of $e^+ e^- \to \tau^+ \tau^-$ events and is assumed to be fully correlated across all tracks. The slow-pion efficiency for momenta less than 0.2 GeV is corrected relative to the tracking efficiency at momenta larger than 0.2 GeV in the laboratory frame. We study $\overline{B}{}^0 \to D^{*+} \pi^-$ decays to determine corrections for three momentum bins spanning $[0.05, 0.12, 0.16, 0.20]$ GeV. The associated statistical and systematic uncertainties of the correction weights affect the  $\mathcal{R}(D^+)$ and $\mathcal{R}(D^{*+})$ and are evaluated using variations of the correction weights, taking into account their correlations. 

A systematic uncertainty also arises from the modeling of the BDT input variables. The most significant mismodeling is observed in the negative region of the $\cos\theta_{BY}$ distribution of the $B_{\mathrm{sig}}$ candidate. To estimate the impact of this mismodeling, MC samples are reweighted to data, with event weights derived from a linear spline fit to the data-to-MC ratio as a function of $\cos\theta_{BY}$. This reweighting is done separately for each of the four categories, yielding distinct spline fits. The event weights are then used to construct a fit, with each template normalized to the nominal fit event count to isolate shape effects. Two approaches were investigated: first, reweighting data while fitting with nominal templates, 
and second, reweighting the fit templates while fitting to nominal data. 
We use the larger shift to assess the uncertainty. 

The uncertainty on the form factors of semitauonic and semileptonic signal are evaluated by using the eigenvariations provided in Ref.~\cite{Bernlochner:2022ywh}. In addition, we assign the difference in central value between the parametrization of  Ref.~\cite{Bernlochner:2022ywh} and Refs.~\cite{Boyd:1995sq,Boyd:1997kz} for semileptonic signal and Ref.~\cite{Caprini:1997mu} for semitauonic signal events, using the information from Refs.~\cite{Fajfer:2012vx,Tanaka:2010se}.

Finally, we assign a 10\% uncertainty to the fraction of continuum events in the background, based on the observed difference in efficiency between off-resonance data and the expected continuum contribution.

\begin{table}[h!]
	\footnotesize
	\renewcommand{\arraystretch}{1.3}
	\begin{center}
		\caption{Systematic uncertainties on $\mathcal{R}(D^+)$ and $\mathcal{R}(D^{*+})$ ranked by the magnitude of the uncertainty on $\mathcal{R}(D^+)$. The percentage values in brackets indicate the relative uncertainty.}
		\begin{tabular}{lll}
			\hline\hline
				Systematic Uncertainty & $\Delta \mathcal{R}(D^+)$ &  $\Delta \mathcal{R}(D^{*+})$ \\
			\hline
			{\bf Additive} & & \\
			$\,$ MC sample size  & $0.033~(7.9\%)$ &  $^{+0.011 (3.4\%)}_{-0.012 (3.9\%)}$ \\
			$\,$ Gap $\cal B$         & $^{+0.014 (3.4\%)}_{-0.038 (9.1\%)} $  & 0.001 (0.1\%) \\
			$$\,$$ LID efficiency ($\mu$)  & $^{+0.025 (6.1\%)}_{-0.016 (3.9\%)}$  & 0.001 (0.1\%) \\
			$$\,$$ Fake rates ($e$)     & $^{+0.016 (3.7\%)}_{-0.008 (2.0\%)}$   & $^0.002~(0.7\%)$ \\
			$\,$ $\pi^\pm$ from $D^* \to D \pi$ & 0.003 (0.7\%)   & 0.001 (0.1\%) \\
			$\,$ Continuum fraction   &  $^{+0.003 (0.7\%)}_{-0.002 (0.4\%)}$   & 0.001 (0.2\%) \\

			$\,$ Gap FFs         & $^{+0.003 (0.6\%)}_{-0.001 (0.3\%)}$  & 0.001 (0.1\%) \\
			$\,$ $\mathcal{B}(\overline{B}{} \to D^{**} \ell \bar \nu_\ell)$ & 0.002 (0.5\%)   & 0.001 (0.1\%) \\
			$\,$ $\overline{B}{} \to D^{**} \ell \bar \nu_\ell$ FFs & 0.001 (0.3\%)   & 0.001 (0.2\%) \\
			$\,$ BDT modeling         & 0.001 (0.3\%)   & 0.001 (0.2\%) \\
            $\,$ $\overline{B}{} \to D^{(*)} \ell \bar \nu_\ell$ / $\tau \bar \nu_\tau$ FFs &$0.001 (0.1\%)$   &$0.001 (0.3\%)$  \\
			$$\,$$ LID efficiency ($e$)  & 0.001 (0.1\%)   & 0.001 (0.3\%) \\
			$$\,$$ Fake rates ($\mu$)     & 0.001 (0.1\%)   & 0.001 (0.1\%) \\
			
			\hline
			{\bf Total Additive Uncertainty} & $^{+0.047 (11.3\%)}_{-0.054 (12.9\%)}$  &  $^{+0.011 (3.6\%)}_{-0.012 (4.0\%)}$  \\
			\hline

			{\bf Multiplicative} & & \\
			$\,$ $\overline{B}{} \to D^{(*)} \ell \bar \nu_\ell$ / $\tau \bar \nu_\tau$ FFs & 0.009 (2.1\%)   & 0.011 (3.5\%) \\
			$\,$ MC sample size  & 0.007 (1.7\%)   & 0.004 (1.2\%) \\
			$$\,$$ LID efficiency ($e$)  & 0.001 (0.2\%)   & 0.001 (0.2\%) \\
			
			$\,$ $\mathcal{B}(\tau^- \to \ell^- \overline \nu_\ell \nu_\tau)$        & 0.001 (0.2\%)   & 0.001 (0.2\%) \\
			$$\,$$ LID efficiency ($\mu$)  & 0.001 (0.1\%)   & 0.001 (0.1\%) \\
			$\,$ Tracking efficiency & 0.001 (0.1\%)  & 0.001 (0.1\%) \\
			$\, $ $\pi^\pm$ from $D^* \to D \pi$ & -- $\quad\,$ (--)  & 0.001 (0.2\%) \\
			\hline
		      {\bf Total Multiplicative Uncertainty} & 0.012 (2.8\%) & 0.011 (3.7\%) \\
			\hline
			{\bf Total Syst. Uncertainty} & $^{+0.049 (11.6\%)}_{-0.056 (13.4\%)}$  &$^{+0.016 (5.2\%)}_{-0.018 (5.8\%)}$ \\
			\hline \hline
			{\bf Total Stat. Uncertainty} & $^{+0.075 (17.9\%)}_{-0.073 (17.5\%)}$  &$^{+0.035 (11.4\%)}_{-0.033 (10.8\%)}$ \\
			\hline
			{\bf Total Uncertainty} & $^{+0.089 (21.4\%)}_{-0.092 (22.0\%)}$  & $^{+0.038 (12.6\%)}_{-0.037 (12.2\%)}$ \\
			\hline\hline
		\end{tabular}
		\label{tab:systematics}
	\end{center}
	\end{table}

%%%%%%%%%%%%%%%%%%%%%%%%%%%%%%%%%%%%%%%%%%%%%%%%%%%%%%%%%%%%%%%%%%%%%%%%%%%%%%%%
\section{Results}\label{sec:results}
%%%%%%%%%%%%%%%%%%%%%%%%%%%%%%%%%%%%%%%%%%%%%%%%%%%%%%%%%%%%%%%%%%%%%%%%%%%%%%%%

\subsection{ $\mathcal{R}(D^+)$ and $\mathcal{R}(D^{+*})$ }

\begin{figure*}[t]
	\centering
        \includegraphics[width=0.6\textwidth]{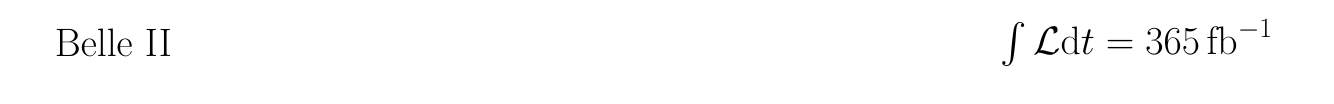} \\
        \vspace{-0.8ex}
        \includegraphics[width=0.4\textwidth,trim=0cm 0cm 0cm 0cm,% 1.05cm, 
        clip]{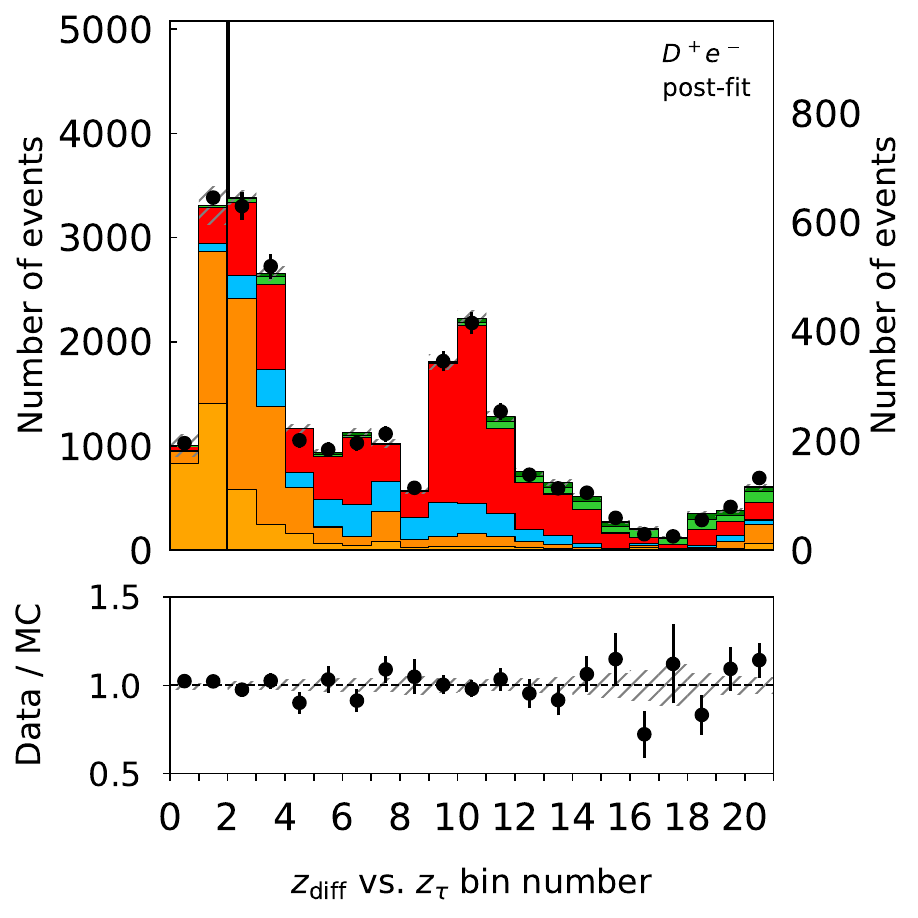} 
        \includegraphics[width=0.4\textwidth,trim=0cm 0cm 0cm 0cm,% 1.05cm, 
        clip]{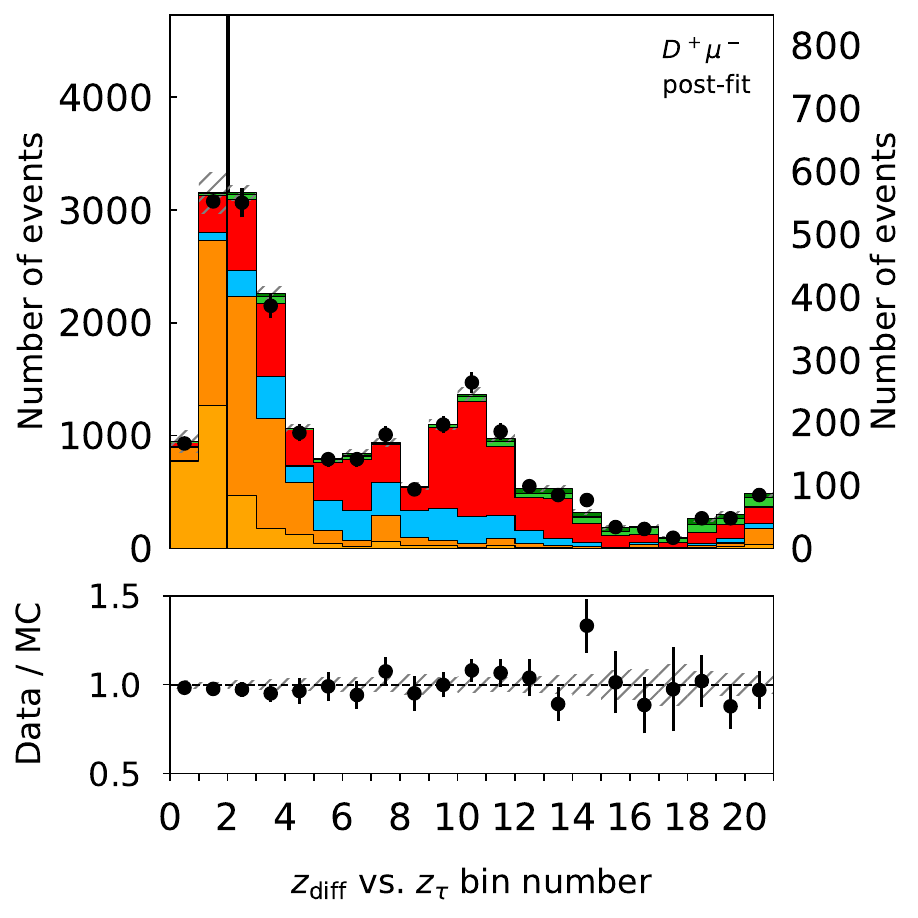}
         \put(-143.,112.){\includegraphics[width=0.19\linewidth,trim=2cm 2cm 2cm 2cm,clip]{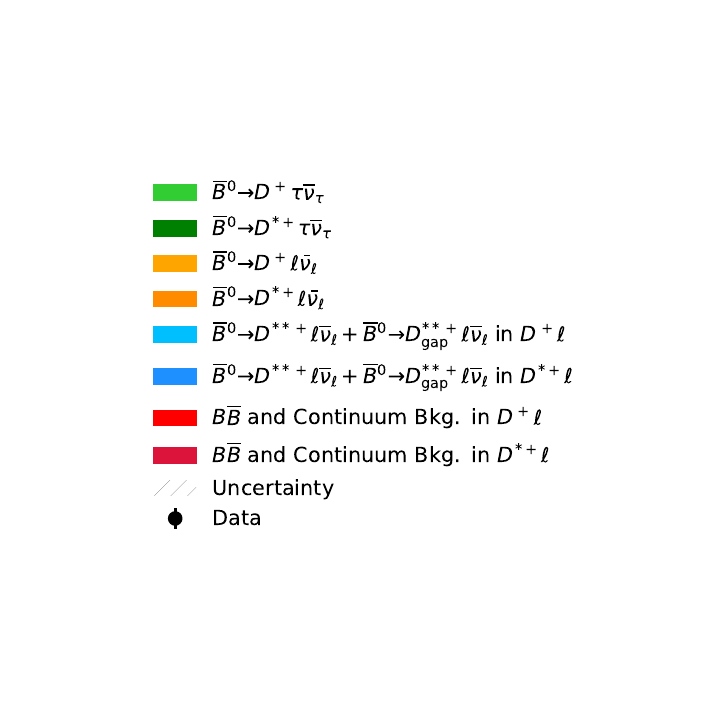}} \quad        
          \\
          \vspace{1ex}
        \includegraphics[width=0.4\textwidth,trim=0cm 0cm 0cm 0cm,% 1.05cm, 
        clip]{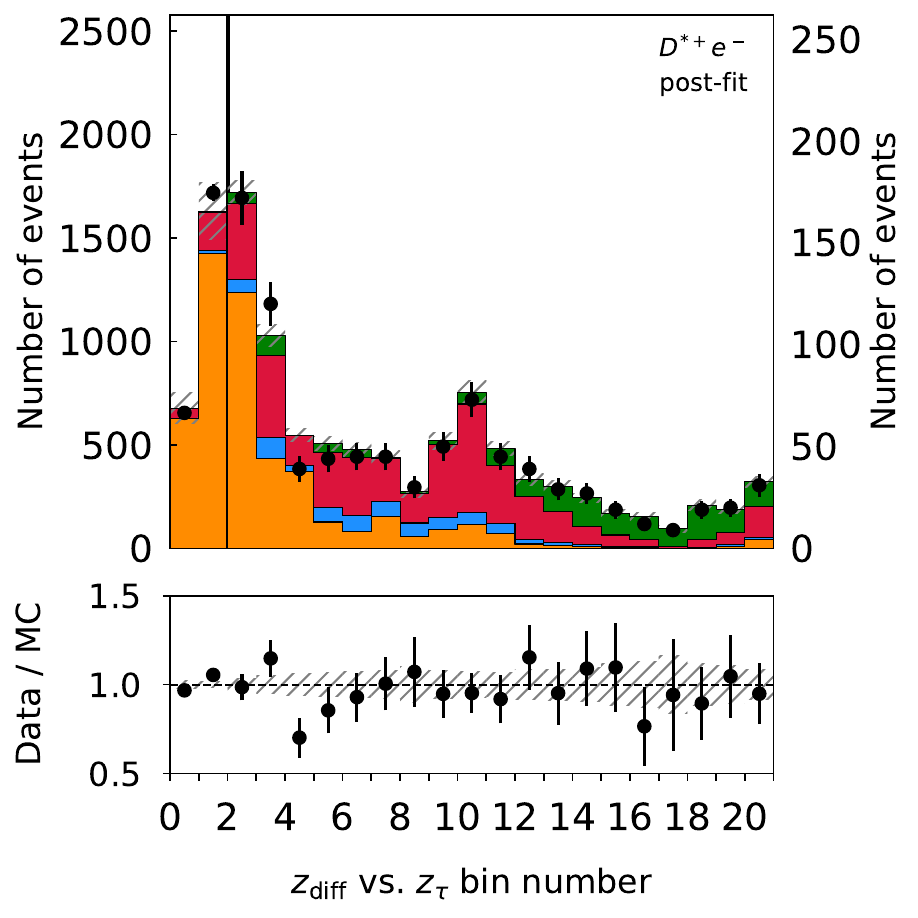}  \quad
        \includegraphics[width=0.4\textwidth,trim=0cm 0cm 0cm 0cm,% 1.05cm, 
        clip]{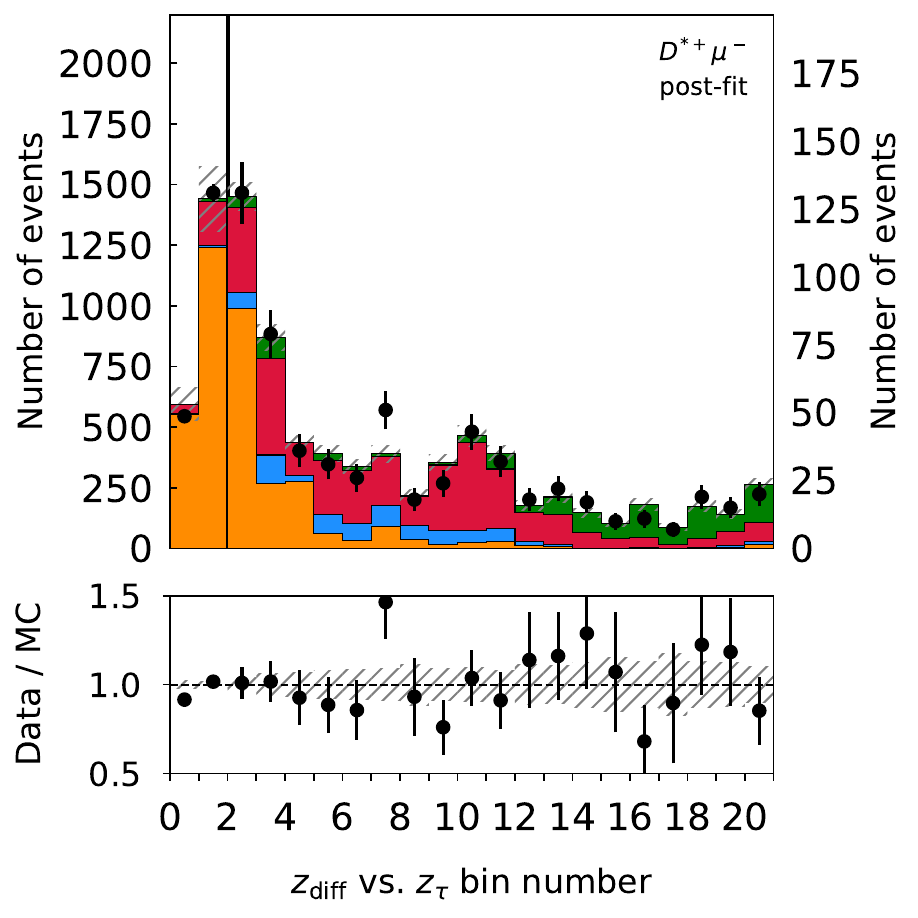} 
         
        \caption{ The fitted classifier distributions for the $D^{+}e^-$, $D^{+}\mu^-$, $D^{*+}e^-$, and $D^{*+}\mu^-$  categories are shown. The hatched regions of the histograms correspond to the systematic uncertainties.
         The black lines demarcate the first two bins from the remaining bins, which contain significantly fewer events, and are thus displayed with two different $y$-axis scales on the left and right sides.
        }
        \label{fig:fit_res}        
\end{figure*}

\begin{figure}[t] 
	\centering
\includegraphics[width=0.52\textwidth]{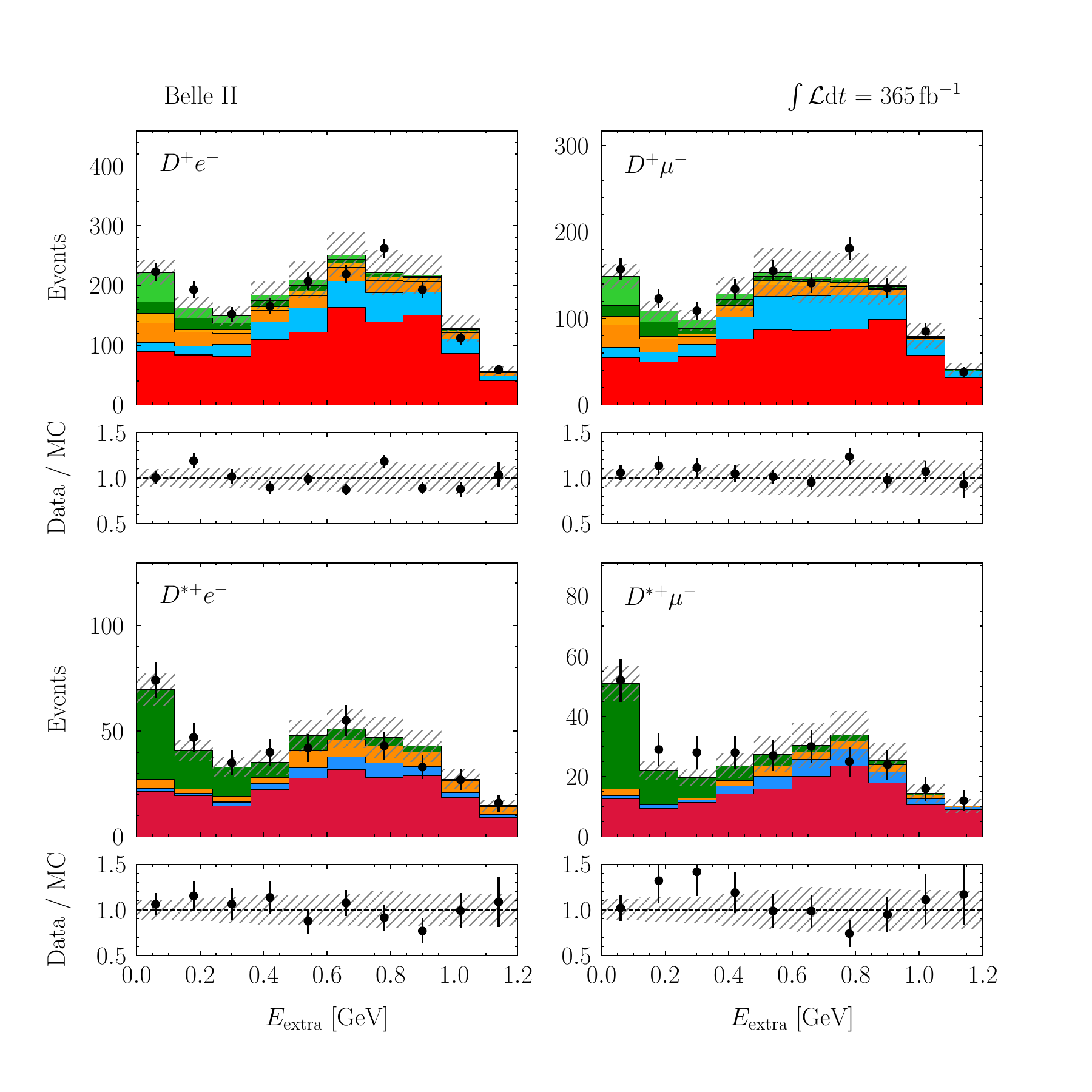}
         \put(-97.,78.){\includegraphics[width=0.23\linewidth,trim=2cm 2cm 2cm 2cm,clip]{legend_only}} \quad        
\caption{ Distributions of the signal enriched bins  (5,6,10--20) for $E_\mathrm{extra}$ with the results of the fit superimposed. }
	\label{fig:postfitEextra_sig_enhanced}
\end{figure}

\begin{figure}[t] 
 	\centering
	\qquad \includegraphics[width=0.52\textwidth]{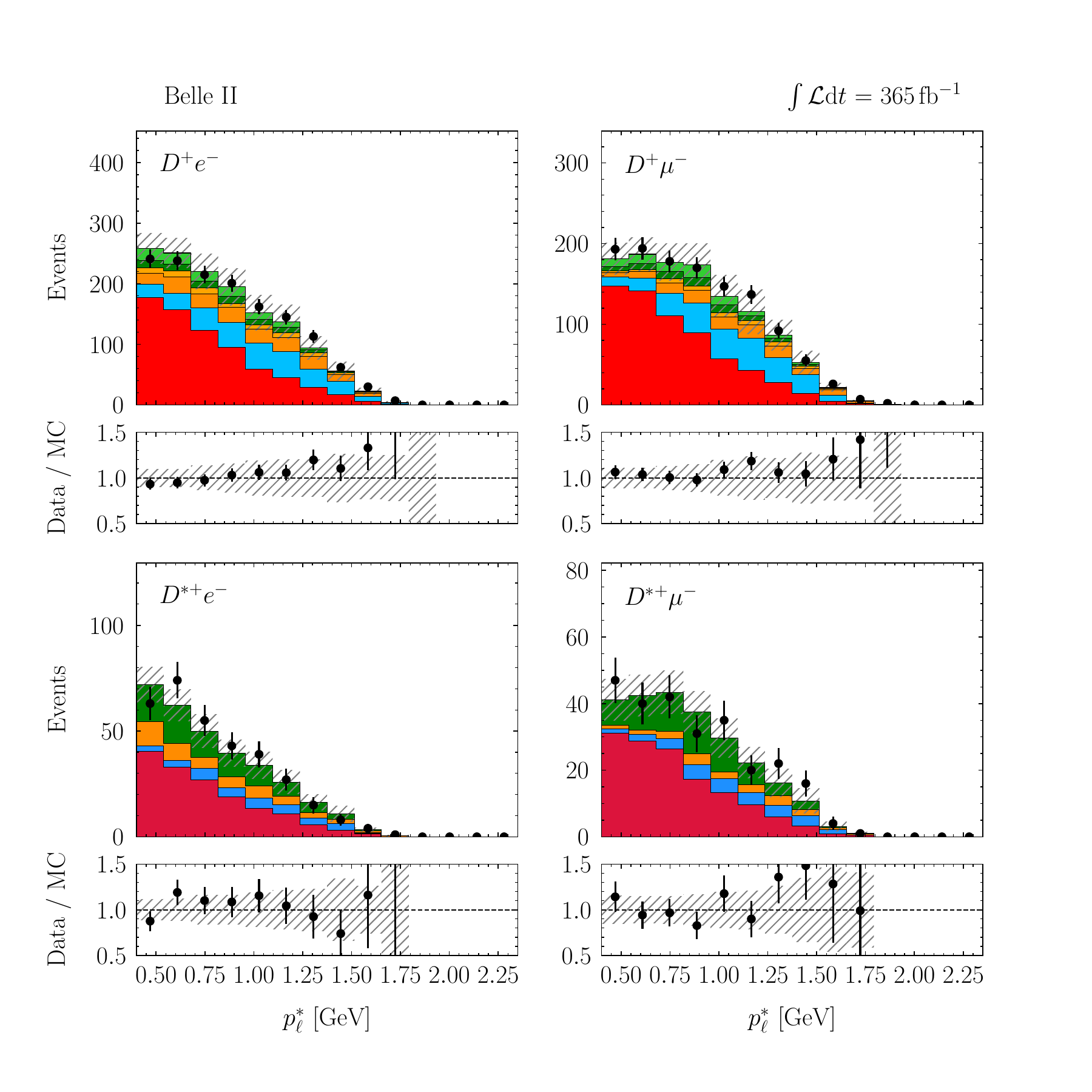}	
         \put(-84.,78.){\includegraphics[width=0.23\linewidth,trim=2cm 2cm 2cm 2cm,clip]{legend_only}} \quad            
 	\caption{ Distributions of the signal enriched bins  (5,6,10--20) for $p_{\ell}^{*}$ with the results of the fit superimposed. }
    \label{fig:postfitpl_sig_enhanced}	
\end{figure}

Fig.~\ref{fig:fit_res} shows the fitted classifier bins for the $D^{(*)+}e^-$ and $D^{(*)+}\mu^-$ categories. We measure 

\begin{align}
  \RDmeas \, , \label{eq:RD} \\
  \RDsmeas \,  \label{eq:RDs} ,
\end{align}
with a correlation of $\rho = -0.24$. These values are compatible with the SM predictions of $\mathcal{R}(D) = 0.296 \pm 0.004$ and $\mathcal{R}(D^*) = 0.254 \pm 0.005$~\cite{Bigi:2016mdz,Gambino:2019sif,Bordone:2019vic,Martinelli:2021onb,Bernlochner:2022ywh,Ray2024,Aoki2022,PhysRevLett.123.091801,Martinelli:2024epjc} within 
1.7 standard deviations. 
The p-value of the fit is 8.3\% and evaluated using the saturated likelihood method~\cite{Baker:1983tu}. 
Figures~\ref{fig:postfitEextra_sig_enhanced}--\ref{fig:postfitpl_sig_enhanced} show the distributions of $E_\mathrm{extra}$ and $p_{\ell}^*$ for the signal enriched bins of the two-dimensional classifier with the postfit scaling applied. More details can be found in Appendix~\ref{app:B}.

\subsection{LFU tests of electrons versus muons }

For the ratio of the semileptonic signal branching fractions of electrons to muons we find,
\begin{align} 
 \mathcal{R}(D^+_{e/\mu}) &= 1.068 ^{+ 0.050}_{-0.048}\text{(stat)} ^{+ 0.024}_{ -0.023}\text{(syst)}  \, ,  \\
 \mathcal{R}(D^{+*}_{e/\mu}) &= 1.079 ^{+0.037}_{- 0.036} \text{(stat)} ^{+ 0.020}_{ -0.019}\text{(syst)} \, ,
\end{align}
with a correlation coefficient of $\rho = -0.398$, both consistent with the expectation of LFU within 1.2 and 1.6 standard deviations, respectively.

\subsection{Consistency checks}

\begin{figure}[!] 
        \hspace{-5.5ex}
        \includegraphics[width=0.45\textwidth,trim=0cm 2cm 0cm 4.5cm, clip]{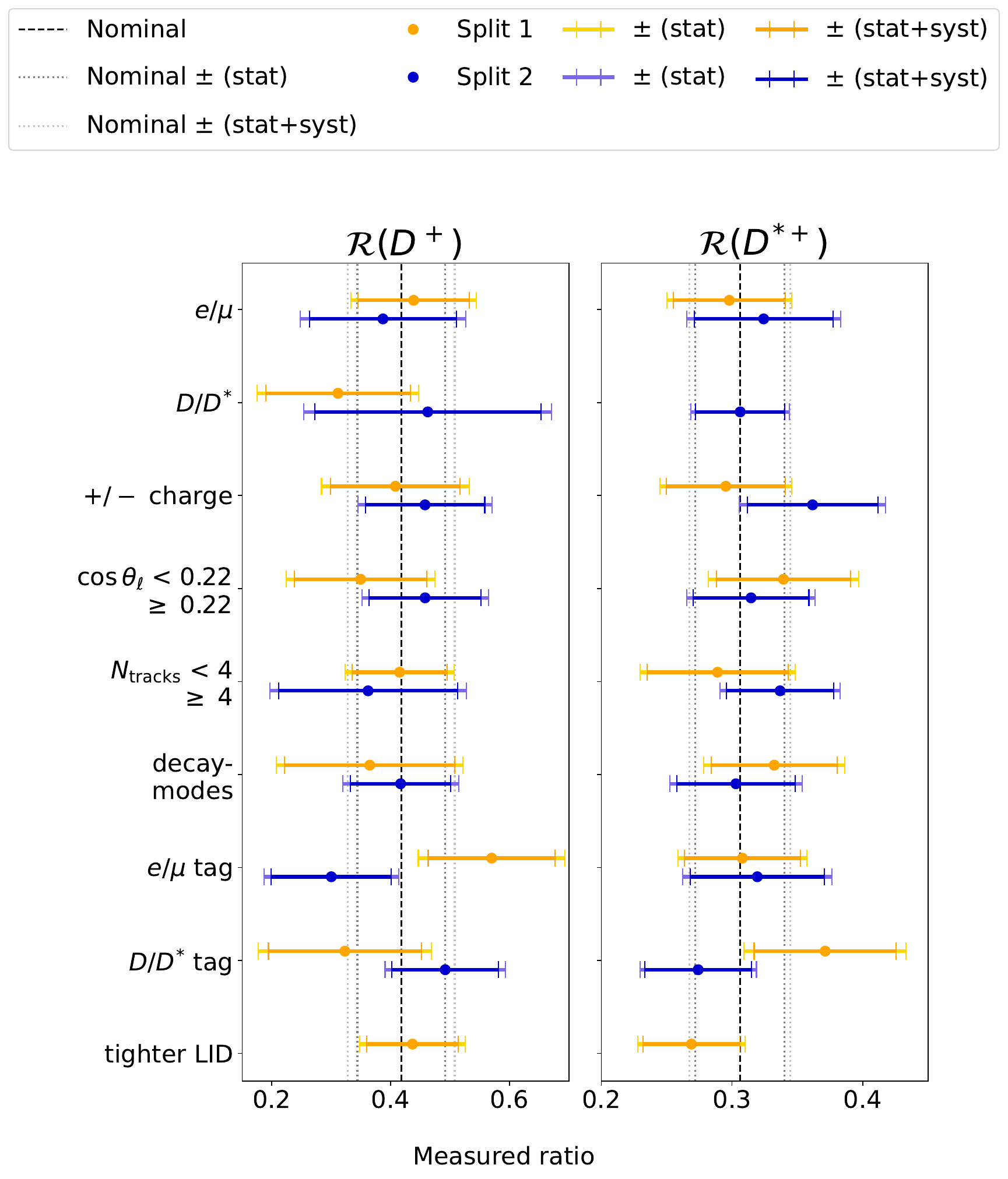} 
        \caption{Summary of the various consistency checks. The central values of each sample split fit (blue and orange error bars), along with the sizes of the statistical and total uncertainties are shown (inner and outer ticks). The nominal results (Eqs.~\eqref{eq:RD} and \eqref{eq:RDs}) are also shown (dashed black line) with statistical and total uncertainty (inner and outer gray line). The left column shows $\mathcal{R}(D)$, and the right column $\mathcal{R}(D^*)$.}
        \label{fig:checks}        
\end{figure}

To assess the stability of the result, we determine $\mathcal{R}(D^{(*)+})$ for various subsamples, fitting them simultaneously to account for correlations in common systematic uncertainties. The tests are summarized in Fig.~\ref{fig:checks}. We consider several sample divisions: First, we split by the lepton flavor of the semitauonic and semileptonic signal, fitting $\mathcal{R}(D^{(*)})$ separately in the reconstructed electron and muon channels. We also decouple the $D^+$ and $D^{*+}$ fit channels, allowing for an independent extraction of $\mathcal{R}(D^+)$ and $\mathcal{R}(D^{*+})$ in the $D^+$ sample, while fitting only $\mathcal{R}(D^{*+})$ in the $D^{*+}$ sample. 

The limited number of $D^*$ feed-down signal events results in a large anticorrelation. 
The obtained $\mathcal{R}(D^+)$ ($\mathcal{R}(D^{+*})$) values in the split datasets are consistent with the nominal result with p-values ranging from 10\%--78\% (20\%--88\%).

Additional cross-checks involve splitting the data by the charge of the signal lepton and by its polar angle, where the latter is divided at $\cos\theta_\ell = 0.22$ to approximately halve the dataset. We also consider a separation based on the number of reconstructed tracks on the signal side (excluding the lepton). 
Here, we choose a threshold of four tracks to ensure a balanced division of the dataset.

Beyond that, the dataset is split according to different $D$ meson reconstruction modes, selecting alternating modes based on their branching fractions. 
Furthermore, we divide the dataset based on the decay modes of the tag-side $B^0$: We partition the dataset by choosing only tag-side semileptonic decays with electrons and with muons, respectively. Second, we further categorize events by distinguishing between tag-side decays with $D$ mesons and those with $D^*$ mesons.
We also test the stability using a more stringent LID selection and find good agreement between the results.

The ratio of branching fractions of $\overline{B}{}^0 \to D^{+} \ell^- \overline \nu_\ell$ over $\overline{B}{}^0 \to D^{*+} \ell^- \overline \nu_\ell$ is $0.40 \pm 0.01$, which is consistent with the prediction from Ref.~\cite{Bernlochner:2022ywh} of $0.417 \pm 0.012$ within 1.1 standard deviations and the world average of Ref.~\cite{HeavyFlavorAveragingGroupHFLAV:2024ctg} of $0.431 \pm 0.014$ within 1.8 standard deviations.

%%%%%%%%%%%%%%%%%%%%%%%%%%%%%%%%%%%%%%%%%%%%%%%%%%%%%%%%%%%%%%%%%%%%%%%%%%%%%%%%
\section{Conclusions}\label{sec:conclusion}
%%%%%%%%%%%%%%%%%%%%%%%%%%%%%%%%%%%%%%%%%%%%%%%%%%%%%%%%%%%%%%%%%%%%%%%%%%%%%%%%

\begin{figure}[t]
	\centering
        \includegraphics[width=0.52\textwidth,trim=0cm 0cm 0cm 0cm, clip]{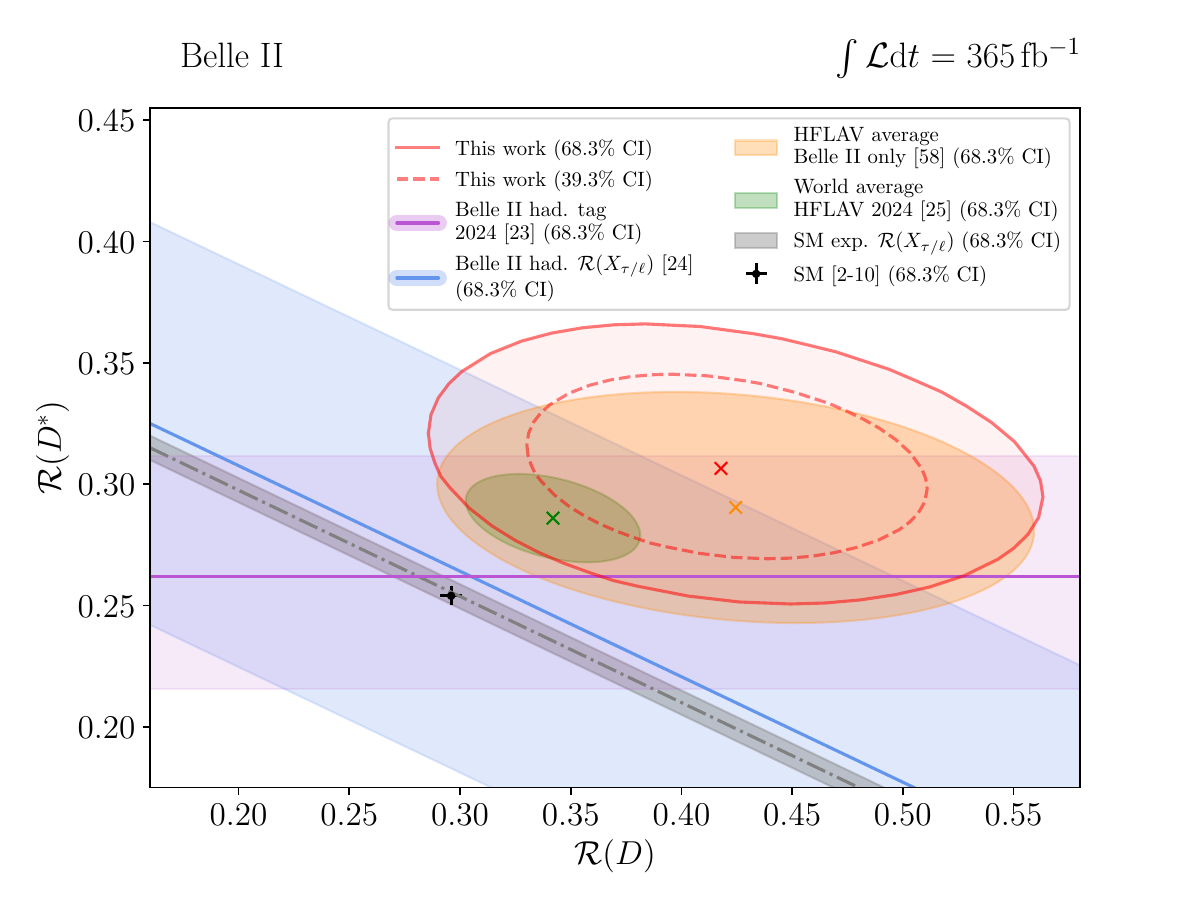} 
        \caption{The measured $\mathcal{R}(D^+)$ and $\mathcal{R}(D^{+*})$ ratios (red marker) with their corresponding 68.3\% CI (red solid contour) as well as 39.3\% CI (red dashed contour), are compared to the Standard Model prediction (black marker) and the world average from Ref.~\cite{HeavyFlavorAveragingGroupHFLAV:2024ctg} (green ellipse). Also displayed are the Belle~II $\mathcal{R}(D^{*})$ measurement using hadronic modes to reconstruct $B_{\mathrm{tag}}$ candidates (dark violet band) and the constraint on $\mathcal{R}(D)$ and $\mathcal{R}(D^{*})$ from the inclusive ${\cal R}(X_{\tau/\ell})$ analysis (blue band).
        An average of the two exclusive Belle~II $\mathcal{R}(D^{(*)})$ measurements (this analysis and the hadronic-tag result), as reported by HFLAV~\cite{HFLAV2025}, is shown in orange. The gray band shows the SM expectation for $\mathcal{R}(X_{\tau/\ell})$, obtained by subtracting non-$\overline{B} \to D^{(*)}(\ell^-/\tau^-)\,\overline{\nu}$ components~\cite{Belle-II:2023aih}. The world average includes only the hadronic-tag Belle~II measurement, excluding both the inclusive ${\cal R}(X_{\tau/\ell})$ result and the current measurement, to preserve a meaningful compatibility assessment of the present result with the existing world average.
}
        \label{fig:RD_RDs}        
\end{figure}

We report measurements of the ratios $\mathcal{R}(D^+)$ and  $\mathcal{R}(D^{*+})$ and test the predictions of lepton-flavor-universality of the SM. For this we analyzed a  \lumi\ data sample, recorded by the Belle~II experiment from 2019--2022. Signal events are selected by first reconstructing the companion $B$ meson in semileptonic modes using a hierarchical approach. The signal side is analyzed using a multiclass multivariate approach, combining the discriminating power of five variables.  The selected events are analyzed using a likelihood fit and we determine 
\begin{align}
  \RDmeas \, , \nonumber \\
  \RDsmeas \, \nonumber ,
\end{align}
 consistent with the SM expectation within 1.7 standard deviations. Fig.~\ref{fig:RD_RDs} shows the 2D confidence intervals (CI) and compares this result with the SM expectation and the Belle~II measurements Refs.~\cite{PhysRevD.110.072020,Belle-II:2023aih}, which analyzed an orthogonal dataset. Further we present the world average from Ref.~\cite{HeavyFlavorAveragingGroupHFLAV:2024ctg} and find our results to be consistent with it within 0.6 standard deviations. The uncertainties on the measurements of the ratios are dominated by statistical uncertainties and the largest systematic uncertainty is the limited simulated sample size used to determine efficiencies, train the multiclass classification algorithm, and determine template shapes.  

%%%%%%%%%%%%%%%%%%%%%%%%%%%%%%%%%%%%%%%%%%%%%%%%%%%%%%%%%%%%%%%%%%%%%%%%%%%%%%%%
\begin{acknowledgments}
%%%%%%%%%%%%%%%%%%%%%%%%%%%%%%%%%%%%%%%%%%%%%%%%%%%%%%%%%%%%%%%%%%%%%%%%%%%%%%%%

\input{acknowledgements-b2.tex}

\end{acknowledgments}

\section{Data
availability}
The data that support the findings of this article are not publicly available upon publication because it is not technically feasible and/or the cost of preparing, depositing, and hosting the data would be prohibitive within the terms of this research project. The data are available from the authors upon reasonable request.

%%%%%%%%%%%%%%%%%%%%%%%%%%%%%%%%%%%%%%%%%%%%%%%%%%%%%%%%%%%%%%%%%%%%%%%%%%%%%%%%
\bibliographystyle{apsrev4-1}
\bibliography{paper}
%%%%%%%%%%%%%%%%%%%%%%%%%%%%%%%%%%%%%%%%%%%%%%%%%%%%%%%%%%%%%%%%%%%%%%%%%%%%%%%%
\onecolumngrid
\clearpage

\FloatBarrier

\begin{appendix}

\section{Selections}\label{app:sel}
Tables~\ref{tab:selections} and ~\ref{tab:Dmodes} summarize the differences in the tag-side and signal-side $B^{0}$ reconstruction and the reconstructed $D$ decay modes. In the table, $p_\ell$ refers to the lepton momentum in the laboratory frame; LID refers to the likelihood-based variables used for lepton identification; $E$ denotes the energy of the ECL clusters of the photon ($\gamma$) candidate; $\cos\theta_{\vec{p}, \vec{v}}$ denotes the cosine of the angle between the momentum vector and the vector defined by the reconstructed vertex. In addition, $m_{\pi^0}$, $m_{K_S^0}$, $m_D$, and $m_{D^*}$ are the invariant masses of the reconstructed neutral pion, $K_S^0$, $D$, and $D^*$ mesons, respectively, while $m^{\mathrm{PDG}}$ refers to their masses determined in \cite{ParticleDataGroup:2024}. The parameter $\sigma$ is the width of the $D$ mass peak as determined from a Gaussian fit and $Q$ represents the energy release in the particle decay. The classifier output from the tag-side reconstruction algorithm is denoted by $\mathcal{P}$. 

Table~\ref{tab:effs_App} lists the ratios of efficiencies of semitauonic and semileptonic events for the four categories. 

\begin{table}[h]
\footnotesize
\begin{center}
\def~{\hphantom{0}}
\caption{Comparison of the selection criteria for $B_{\mathrm{sig}}$ and $B_{\mathrm{tag}}$ reconstruction.}
\vspace{0.5ex}
\begin{tabular}{lll}
\toprule
\textbf{Selection} & \textbf{$B_{\mathrm{tag}}$} & \textbf{$B_{\mathrm{sig}}$} \\
\midrule
\addlinespace
tracks &  & except for $\pi^+$ from $D^{*+} \to D^{0} \pi^+$: \\
& $|d_0| < 2$\,cm, $|z_0| < 4$\,cm & $|d_0| < 0.5$\,cm, $|z_0| < 2.0$\,cm \\
& & within CDC angular acceptance with $\geq 1$ hit \\
\addlinespace
\midrule
\addlinespace
$e$ & & LID$ > 0.5$ \\
& $p_\ell^* > 1$\,GeV & $p_\ell > 0.2$\,GeV \\
\addlinespace
\midrule
\addlinespace
$\mu$ & & LID$ > 0.9$ \\
& $p_\ell^* > 1$\,GeV & $p_\ell > 0.4$\,GeV \\
\addlinespace
\midrule
\addlinespace
$\gamma$ & forward: $E > 0.10$\,GeV & forward: $E > 0.08$\,GeV \\
& barrel: $E > 0.09$\,GeV & barrel: $E > 0.03$\,GeV \\
& backward: $E > 0.16$\,GeV & backward: $E > 0.06$\,GeV \\
& & $> 1$ crystal \\
& & within CDC angular acceptance \\
& & measured time within $< 200$\,ns of exp. event time \\
& & for $\pi^{0}$ candidates: \\
& & optimized requirement based on the distance to the \\
& & nearest track and a shower-shape classifier \\
\addlinespace
\midrule
\addlinespace
$\pi^0$ & $0.08$\,GeV $< m_{\pi^{0}} < 0.18$\,GeV & $0.12$\,GeV $< m_{\pi^{0}} < 0.145$\,GeV \\
\addlinespace
\midrule
\addlinespace
$K_S^0$ & $0.4$\,GeV$ < m_{K_S^0} < 0.6$\,GeV & $0.45$\,GeV $< m_{K_S^0} < 0.55$\,GeV \\
& & flight distance $> 0$ \\
& & significance of distance $> 0.5$ \\
& & $\cos\theta_{\vec{p}, \vec{v}} > 0.8$ \\
\addlinespace
\midrule
\addlinespace
$D$ & $1.7$\,GeV $< m_D < 1.95$\,GeV  & $m_D^{\mathrm{PDG}} - 2.5\sigma < m_D < m_D^{\mathrm{PDG}} + 2.5\sigma$ \\
 \vspace{0.5mm} & & with $m_{D^0}^{\mathrm{PDG}} = (1.86484 \pm 0.00005)$\,GeV  \\
 \vspace{0.5mm}
& & and $m_{D^+}^{\mathrm{PDG}} = (1.86966 \pm 0.00005)$\,GeV  \\
& & $\sigma(\text{modes with }\pi^+) = 0.0027$\,GeV \\
& & $\sigma(\text{modes with }\pi^0) = 0.005$\,GeV \\
\addlinespace
\midrule
\addlinespace
$D^{*}$ & $0 < Q < 0.3$\,GeV & $0.13$\,GeV $ < m_{D^{*+}} - m_{D^0} < 0.16$\,GeV \\
\addlinespace
\midrule
\addlinespace
$B$ & $1.75 < \cos\theta_{BY} < 1.1$ & $-15 < \cos\theta_{BY} < 1.1$ \\
& $\mathcal{P} > 0.1$ & \\
& cand. with largest $\mathcal{P}$ & cand. with largest $p$-value from $D$ vertex fit \\
\bottomrule
\end{tabular}
\label{tab:selections}
\end{center}
\end{table}

\renewcommand{\arraystretch}{1.3}
\begin{table}[h]
\footnotesize
    \begin{center}
		\def~{\hphantom{0}}
		\caption{Reconstructed $D$ modes used in the reconstruction of $B_{\mathrm{tag}}$ and $B_{\mathrm{sig}}$ candidates.}
		\begin{tabular}{lll}
\hline \hline        
\textbf{Decay mode}  & \textbf{tag side} & \textbf{signal side} \\
\hline
$D^0 \to K^- \pi^+ \pi^0$&$\checkmark$ & $\checkmark$  \\
$D^0 \to K^- \pi^+ \pi^+ \pi^-$&$\checkmark$ &$\checkmark$ \\
$D^0 \to K^- K^+ K_S^0$&$\checkmark$ &$\checkmark$  \\
$D^0 \to K^- K^+$&$\checkmark$ &$\checkmark$ \\
$D^0 \to K^- \pi^+$&$\checkmark$ &$\checkmark$ \\
$D^0 \to K_S^0 \pi^+ \pi^-$&$\checkmark$ &$\checkmark$ \\
$D^0 \to \pi^- \pi^+$&$\checkmark$ & $\checkmark$\\
$D^0 \to K^- \pi^+ \pi^0 \pi^0$&$\checkmark$  & \\
$D^0 \to K^- \pi^+ \pi^+ \pi^- \pi^0$&$\checkmark$  &  \\
$D^0 \to \pi^- \pi^+ \pi^0$&$\checkmark$  & \\
$D^0 \to \pi^- \pi^- \pi^+ \pi^0$&$\checkmark$ & \\
$D^0 \to \pi^- \pi^+ \pi^+ \pi^-$ &$\checkmark$  & \\
$D^0 \to K_S^0 \pi^0$&$\checkmark$  &  \\
$D^0 \to K_S^0 \pi^+ \pi^- \pi^0$&$\checkmark$  &  \\
$D^0 \to K^- K^+ \pi^0$&$\checkmark$  & \\\hline

$D^+ \to K^- \pi^+ \pi^+$ &$\checkmark$& $\checkmark$  \\
$D^+ \to K^0_S \pi^+ \pi^0$&$\checkmark$ & $\checkmark$  \\
 $D^+ \to K^0_S \pi^+ \pi^+ \pi^-$ & $\checkmark$ & $\checkmark$  \\
 $D^+ \to K^0_S \pi^+$ &$\checkmark$& $\checkmark$  \\
 $D^+ \to K^- K^+ \pi^+$ & $\checkmark$& $\checkmark$  \\
 $D^+ \to K^0_S K^+$  & & $\checkmark$  \\
$D^+ \to \pi^+ \pi^0$&$\checkmark$ & \\
$D^+ \to K^- \pi^+ \pi^+ \pi^0$&$\checkmark$ & \\
$D^+ \to \pi^+ \pi^+ \pi^-$&$\checkmark$ & \\
$D^+ \to \pi^+ \pi^+ \pi^- \pi^0$&$\checkmark$ & \\
 $D^+ \to K^+ K_S^0 K_S^0$ &$\checkmark$ & \\
 $D^0 \to K^- K^+ \pi^+ \pi^0$ &$\checkmark$ & \\
			\hline \hline
\end{tabular}
\label{tab:Dmodes}
\end{center}
\end{table}

\begin{table}[h!]
\caption{Ratios of efficiencies between semitauonic and semileptonic signal. The uncertainties correspond to the systematic uncertainties. }
\label{tab:effs_App}
\centering
\renewcommand{\arraystretch}{1.8}
\begin{tabular}{ccccc}
\hline
\hline
\textbf{Category} & $D^+e^-$ & $D^+\mu^-$ & $D^{+*}e^-$ & $D^{*+}\mu^-$ \\
\hline
$  \epsilon_{D^{(*)}\tau} / \epsilon_{D^{(*)}\ell}$ & $0.190 \pm 0.005$ & $0.142 \pm 0.004$ & $0.185 \pm 0.007$ & $0.185 \pm 0.007$ \\
\hline
\hline
\end{tabular}
\end{table}

\FloatBarrier

\clearpage

\section{Fit Details}\label{app:B}
\begin{figure}[!b] 
	\centering
	 \includegraphics[width=0.49\textwidth]{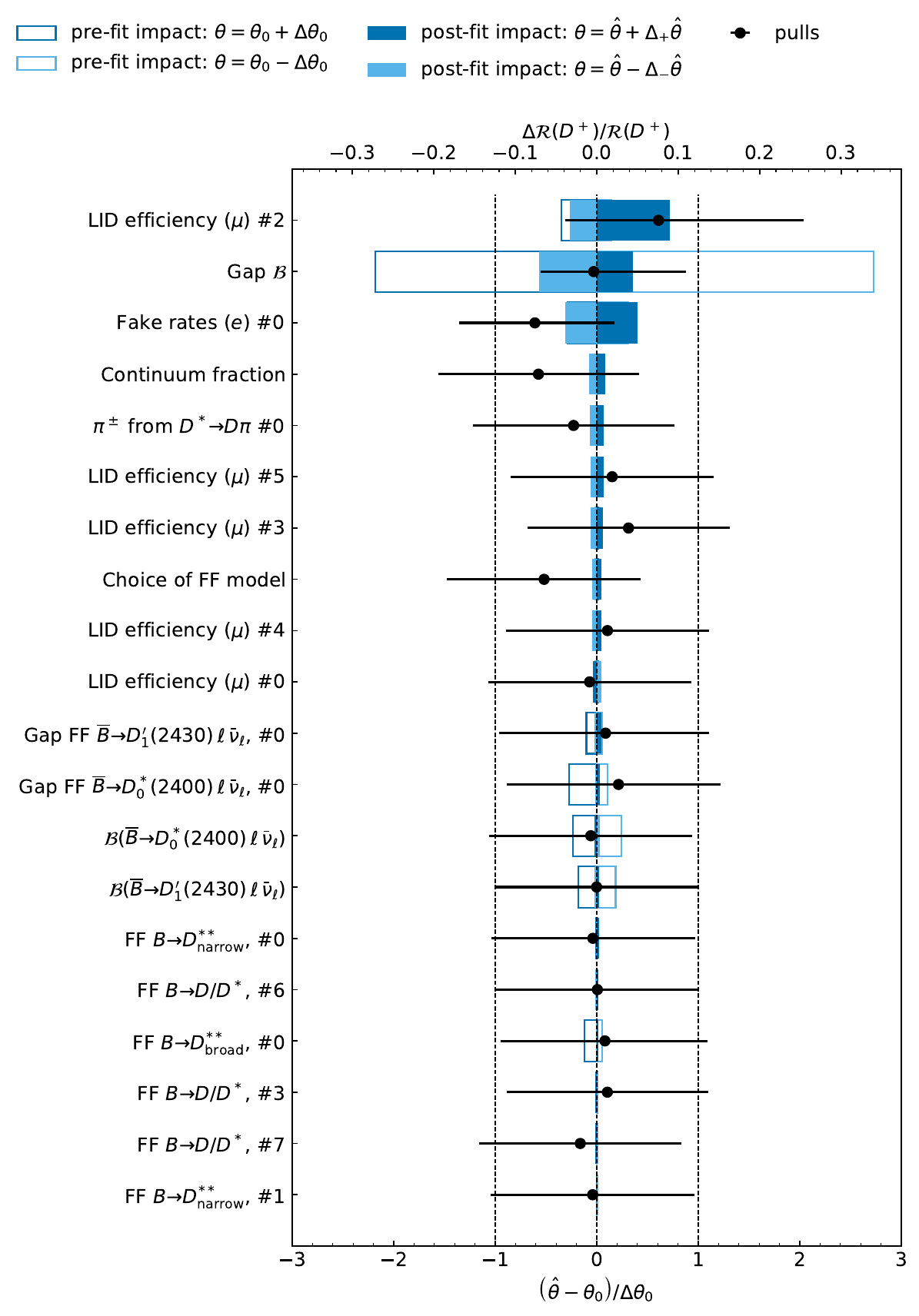}
     \includegraphics[width=0.49\textwidth]{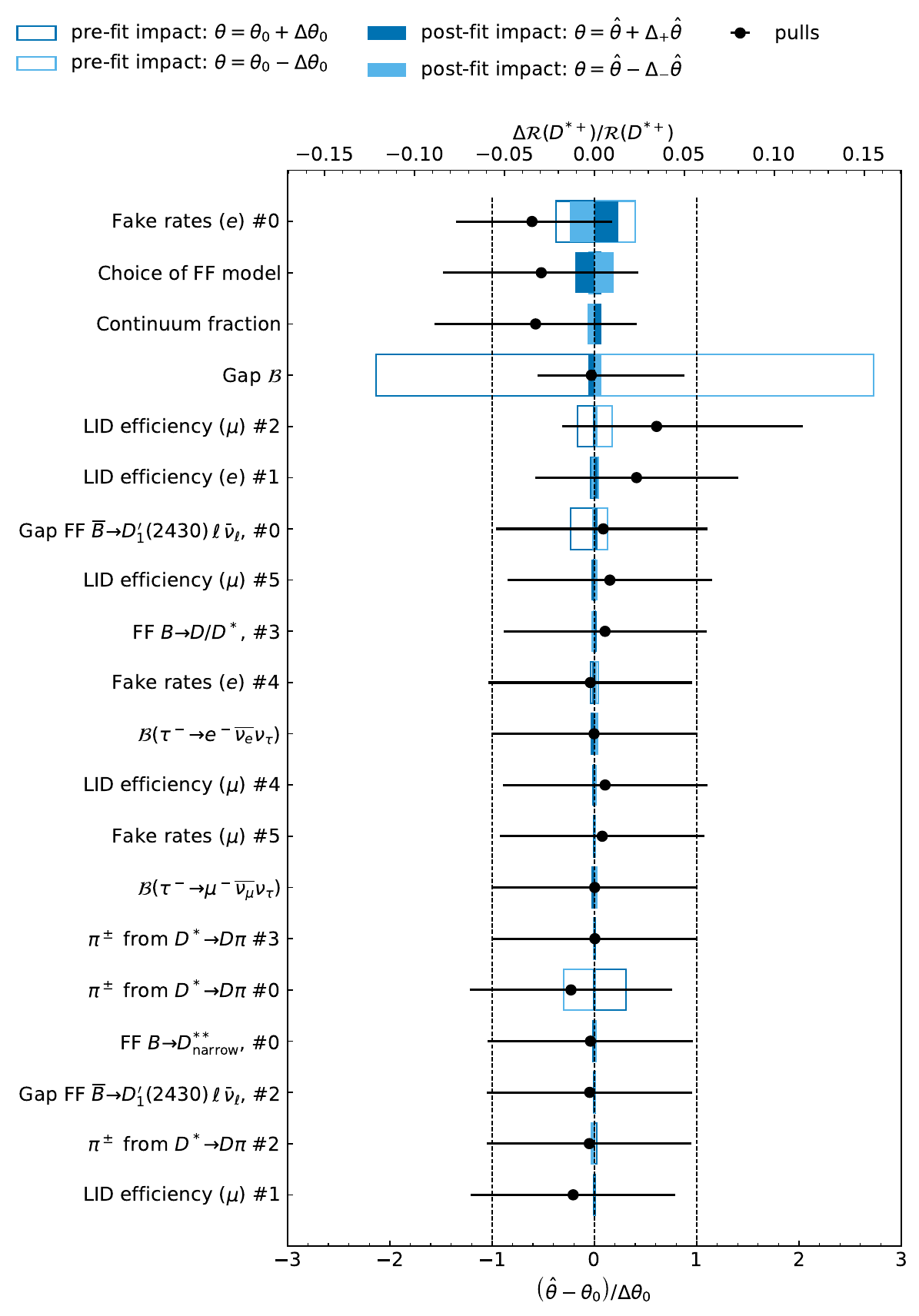}
	\caption{Nuisance parameter ranking for $\mathcal{R}(D^+)$ (left) and  $\mathcal{R}(D^{*+})$ (right) showing the 20 leading nuisance parameters with the largest impact on the respective fitted ratio.}
	\label{fig:pullsRD}
\end{figure}

Fig.~\ref{fig:pullsRD} shows the leading 20 systematic uncertainties on $\cal{R}(D^+)$ and $\cal{R}(D^{*+})$, excluding the dominant systematic uncertainty due to the limited MC sample size. If stated, the number (\#) indicates the eigendirection of the uncertainty source. The uncertainty ``choice of FF model'' parametrizes the impact of changing the semitauonic and semileptonic signal form factor parametrization, cf. Sec.~\ref{sec:systematics}. Within the FF uncertainties for the $D^{**}$ resonances, we distinguish between $D_{\mathrm{broad}}^*$ resonances [$D_0^*(2400)$ and $D_1(2430)$] and $D_{\mathrm{narrow}}^*$ resonances [$D_1(2420)$ and $D_2^*(2460)$] as they are described by different parametrizations due to their varying decay widths.

The pull is defined as the difference of the postfit value ($\widehat \theta$) with respect to the nominal input value of the nuisance parameter ($\theta_0$), and normalized with the postfit uncertainty ($\Delta \theta$). The sizes of the error bars on the pulls correspond to the asymmetric confidence intervals obtained from the likelihood profile.

Also shown are the impacts of the relative uncertainties $\Delta \cal{R}(D^+) / \cal{R}(D^+)$ and $\Delta \cal{R}(D^{*+}) /  \cal{R}(D^{*+})$ on the determined ratios. 
We estimate the prefit impact by analyzing how variations in the NP affect the POI, independent of the other fitted parameters: In order to isolate the specific contribution of the NP, we construct an Asimov dataset in which the NP under investigation is maintained at its initial value, while all other parameters are set to their best-fit values obtained from the nominal fit. This dataset is then refitted, keeping all parameters except the respective POI [$\Delta \cal{R}(D^+) / \cal{R}(D^+)$ or $\Delta \cal{R}(D^{*+}) /  \cal{R}(D^{*+})$] fixed to their best-fit values, 
while shifting the NP to $\theta_0 \pm \Delta \theta_0$, where $ \Delta \theta_0$ denotes its prefit uncertainty. The prefit impact is then quantified as the relative change in the POI compared to a reference POI value, determined in the same way with the NP set at its initial value.

The postfit impact is defined as the relative deviation of the POI from its nominal value, considering only postfit systematic variations. To evaluate this, the best-fit value of the NP from the nominal fit, $\widehat{\theta}$, is used and varied within its estimated uncertainties, $\widehat{\theta} \pm \Delta_{\pm} \widehat{\theta}$, where $\Delta_{\pm} \widehat{\theta}$ represents the asymmetric uncertainty obtained from the likelihood scan. In this case, only the NP under consideration is fixed, while all other parameters are allowed to vary freely during the fit to real data.

Table \ref{tab:postfit_yields} lists the yields within each category as calculated from the fit parameters.

\begin{table}[!htb]
\footnotesize
\caption{Determined yields and uncertainties for all four categories.}
\renewcommand{\arraystretch}{1.3}
\begin{tabular}{l|cccc}
\hline\hline
\textbf{Sample} & $D^+e$ & $D^+\mu$ & $D^{*+}e$ & $D^{*+}\mu$ \\
\hline
$\overline{B}{}^0 \to D^+ \ell \bar{\nu}_{\ell}$ & 2519 $\pm$ 68 & 2233 $\pm$ 61 &  &  \\
$\overline{B}{}^0 \to D^{*+} \ell \bar{\nu}_{\ell}$ & 2486 $\pm$ 63 & 2323 $\pm$ 58 & 2344 $\pm$ 51 & 1961 $\pm$ 44 \\
$\overline{B}{}^0 \to D^+ \tau \bar{\nu}_{\tau}$ & 191 $\pm$ 41 & 155 $\pm$ 65 &  &  \\
$\overline{B}{}^0 \to D^{*+} \tau \nu$ & 106 $\pm$ 14 & 84 $\pm$ 11 & 155 $\pm$ 19 & 111 $\pm$ 14 \\
\hline
$\overline{B}{} \to D^{**} \ell \overline{\nu}_{\ell} / \overline{B} \to D^{**}_{\mathrm{gap}} \ell \overline{\nu}_{\ell}$ & 653 $\pm$ 112 & 586 $\pm$ 102 & 87 $\pm$ 55 & 75 $\pm$ 46 \\
$B \overline{B}{}$ and $\mathrm{Continuum\ Bkg. }$ & 2177 $\pm$ 145 & 1582 $\pm$ 149 & 611 $\pm$ 95 & 497 $\pm$ 83 \\
\hline
Data & 8219 & 6854 & 3241 & 2621 \\
\hline\hline
\end{tabular}
\label{tab:postfit_yields}
\end{table}

\FloatBarrier

\clearpage

\section{Postfit Projections of BDT Input Variables}\label{app:A2}
\begin{figure}[!htbp] 
	\centering
	 \includegraphics[width=0.8\textwidth]{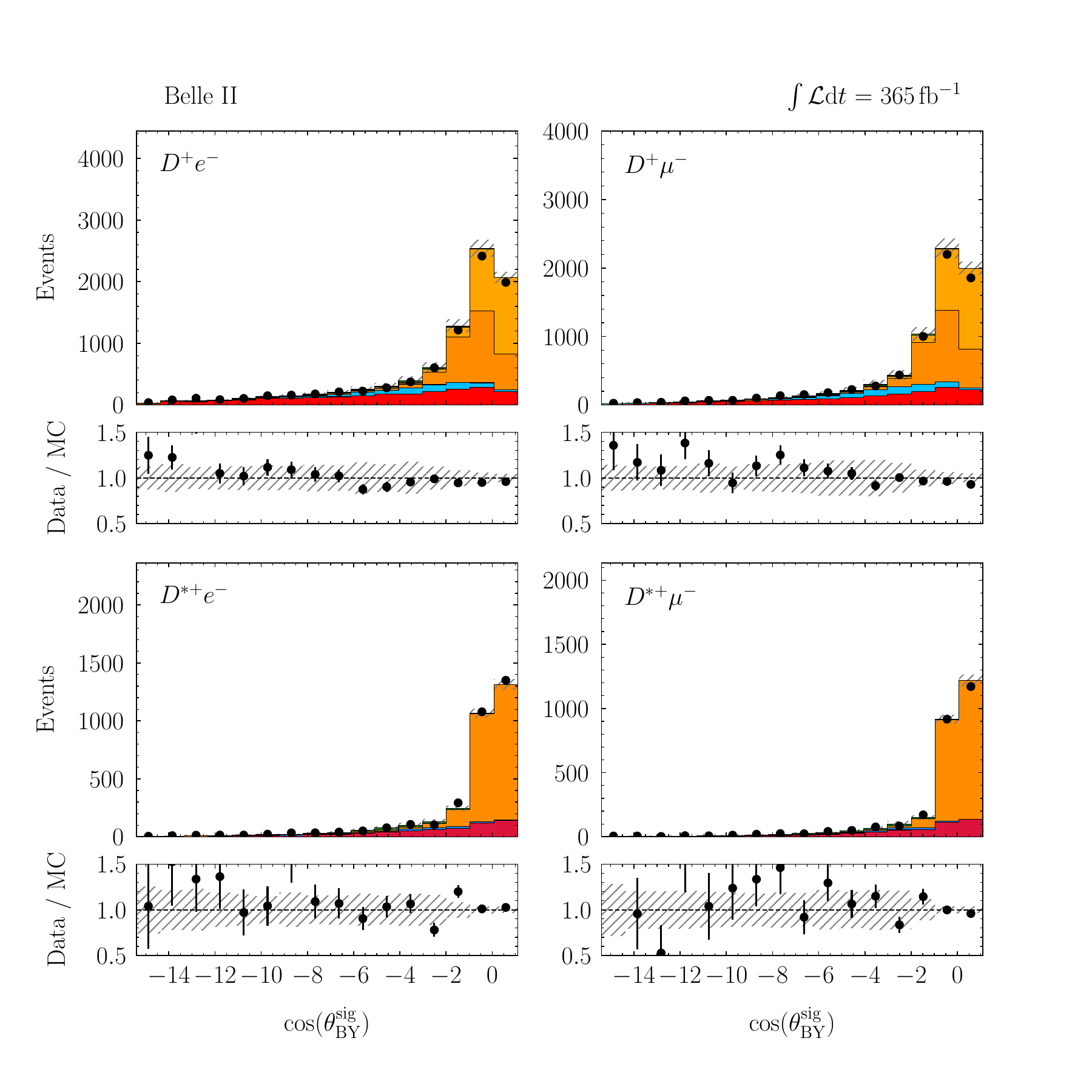}
     \put(-45.,260.){\includegraphics[width=0.25\linewidth,trim=2cm 3cm 2.5cm 3cm, clip]{legend_only}} \quad %
 
	\caption{Distribution of $\cos\theta_{BY}$ with best-fit scaling, including all systematic uncertainties for the four fit categories: $D^{+}e^{-}$, $D^{+}\mu^{-}$, $D^{*+}e^{-}$ and $D^{*+}\mu^{-}$. }
	\label{fig:postfitcosby}
\end{figure}

\begin{figure}[!htbp] 
	\centering
	  \includegraphics[width=0.8\textwidth]{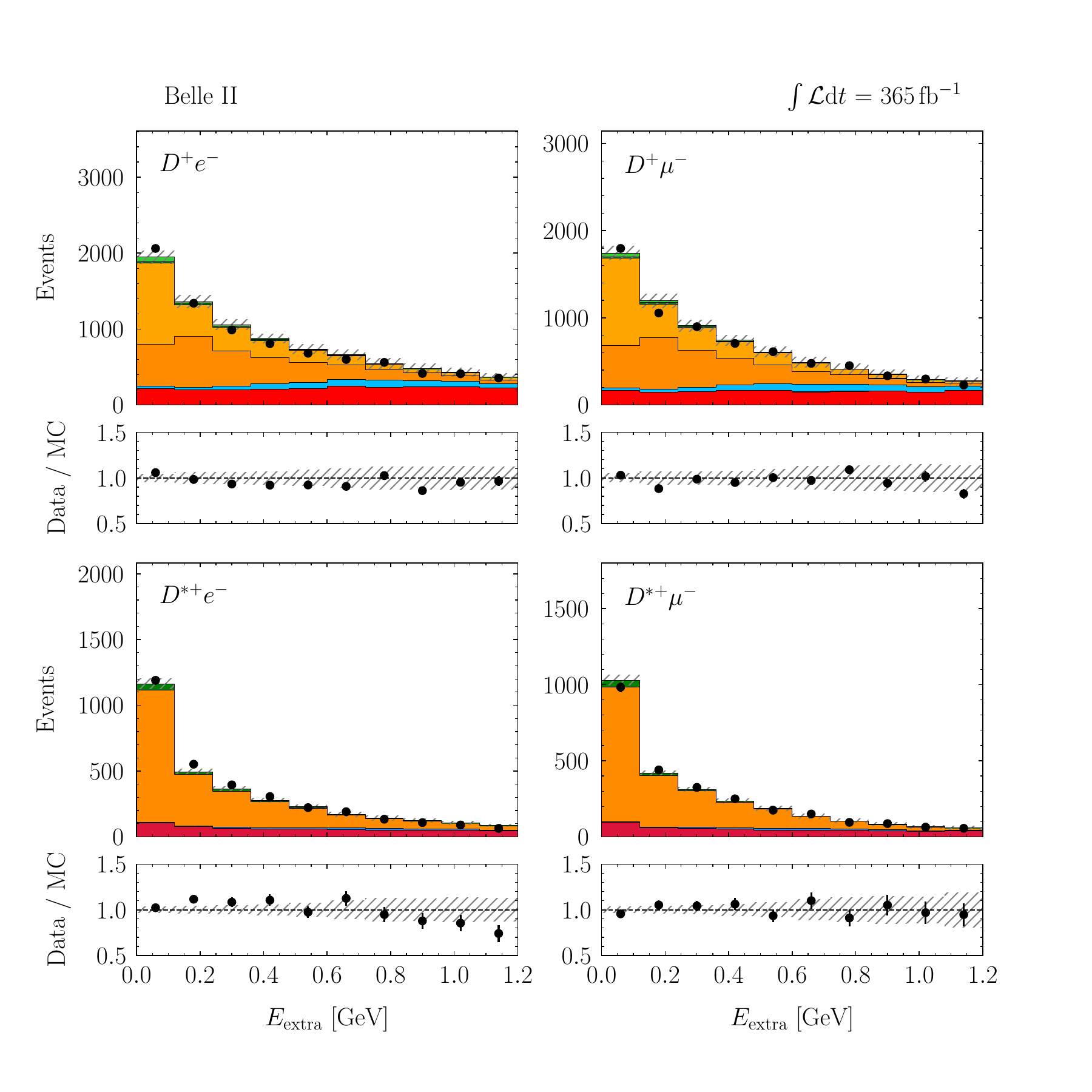}
  \put(-45.,260.){\includegraphics[width=0.25\linewidth,trim=2cm 3cm 2.5cm 3cm, clip]{legend_only}} \quad 
	\caption{Distribution of $E_\mathrm{extra}$ with best-fit scaling, including all systematic uncertainties for the four fit categories.}
	\label{fig:postfiteextra}
\end{figure}

\begin{figure}[!htbp] 
	\centering
    \includegraphics[width=0.8\textwidth]{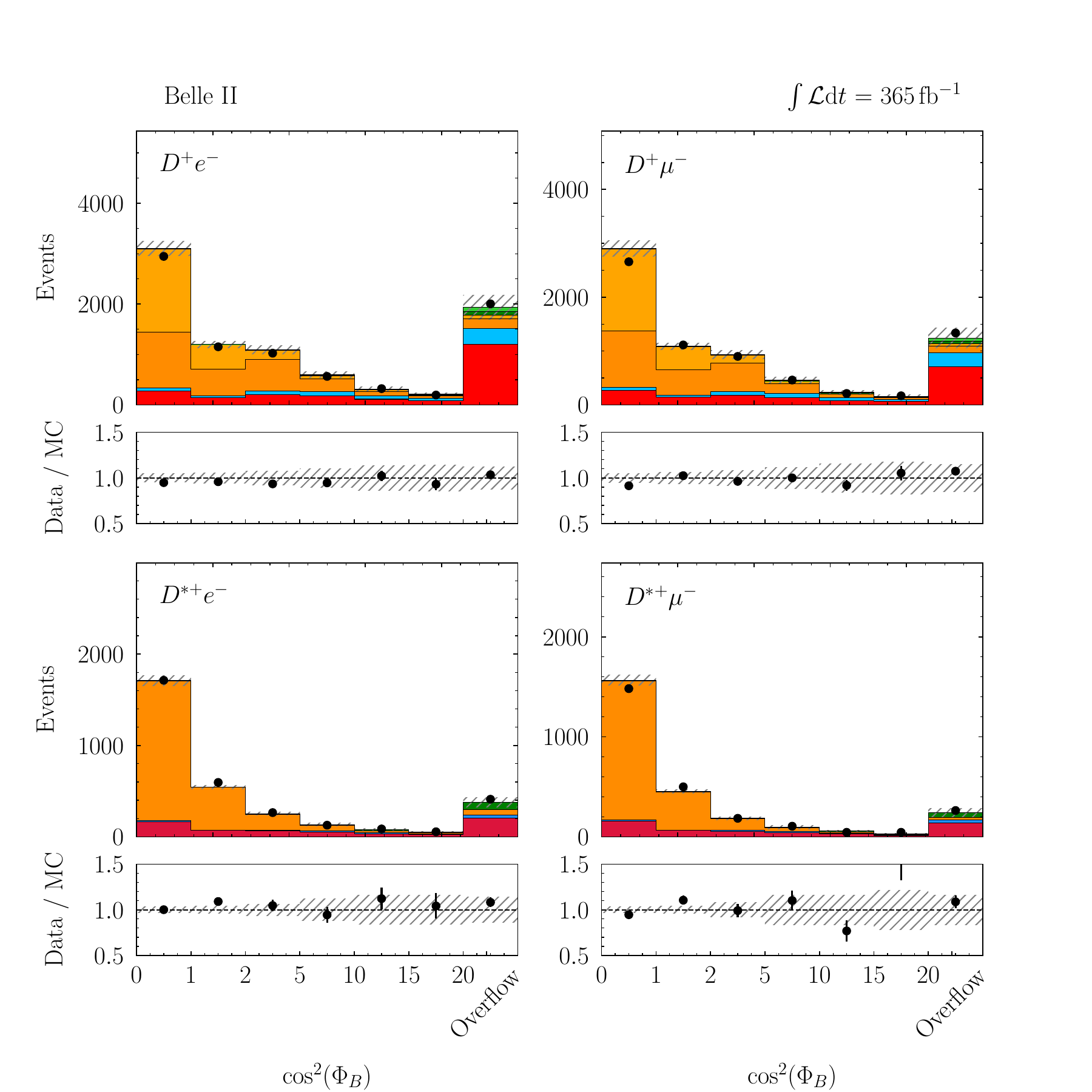}
       \put(-45.,260.){\includegraphics[width=0.25\linewidth,trim=2cm 3cm 2.5cm 3cm, clip]{legend_only}} \quad %
	\caption{Distribution of $\cos^2\Phi_B$ with best-fit scaling, including all systematic uncertainties for the four fit categories. Note the use of nonuniform binning, including a large overflow bin, which accounts for the absence of significant structure in the region of high $\cos^2\Phi_B$.}
	\label{fig:postfitcos2phib}
\end{figure}

\begin{figure}[!htbp] 
	\centering
	 \includegraphics[width=0.8\textwidth]{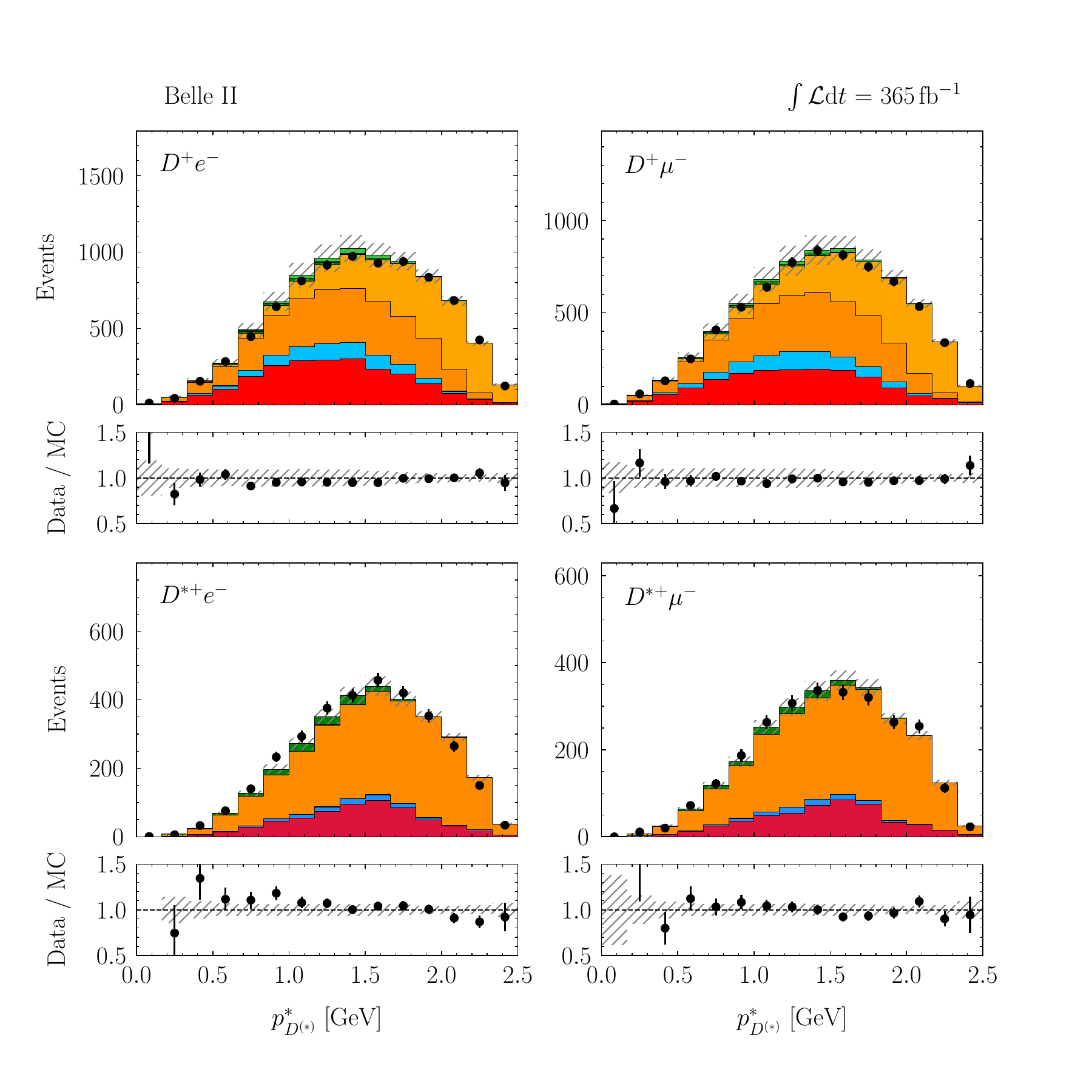}
          \put(-45.,260.){\includegraphics[width=0.25\linewidth,trim=2cm 3cm 2.5cm 3cm, clip]{legend_only}} \quad %

	\caption{Distribution of $p_{D}^{*}$ with best-fit scaling, including all systematic uncertainties for the four fit categories.}
	\label{fig:postfitpD}
\end{figure}

\begin{figure}[!htbp] 
	\centering
	  \includegraphics[width=0.8\textwidth]{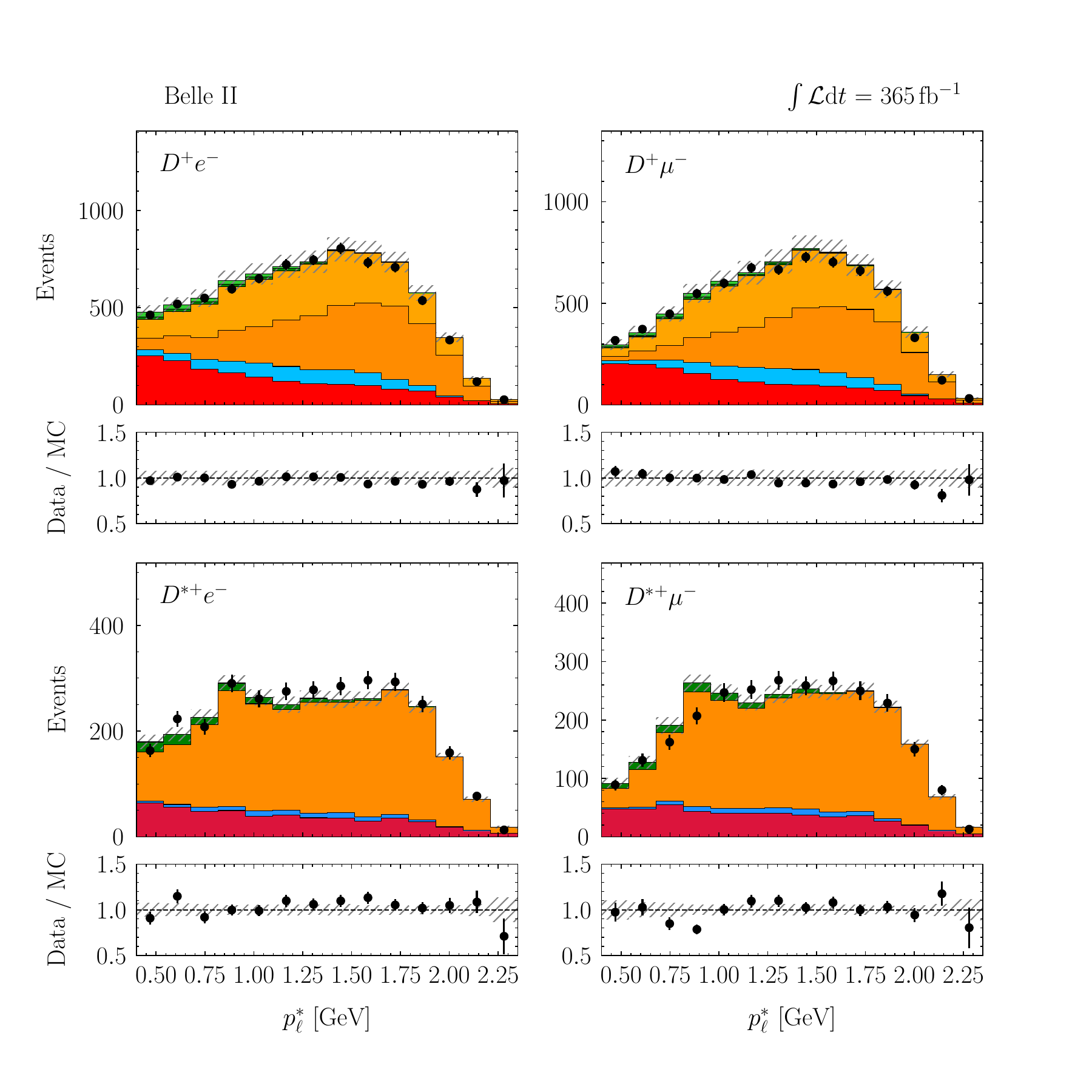} 
        \put(-45.,260.){\includegraphics[width=0.25\linewidth,trim=2cm 3cm 2.5cm 3cm, clip]{legend_only}} \quad %
 \caption{Distribution of $p_{\ell}^{*}$ with best-fit scaling, including all systematic uncertainties for the four fit categories.}
	
	\label{fig:postfitpl}
\end{figure}

Figures~\ref{fig:postfitcosby}--\ref{fig:postfitpl} show the distributions of the input variables utilized in the multivariate algorithm used in the signal classification, arranged by their importance in the classifier training. 

To evaluate the agreement between simulation and data, the fit templates are scaled to their best-fit values.

\FloatBarrier

\section{Systematic uncertainties for the LFU tests $e$ vs. $\mu$}\label{app:A3}
\begin{figure}[!b] 
	\centering
	 \includegraphics[width=0.49\textwidth]{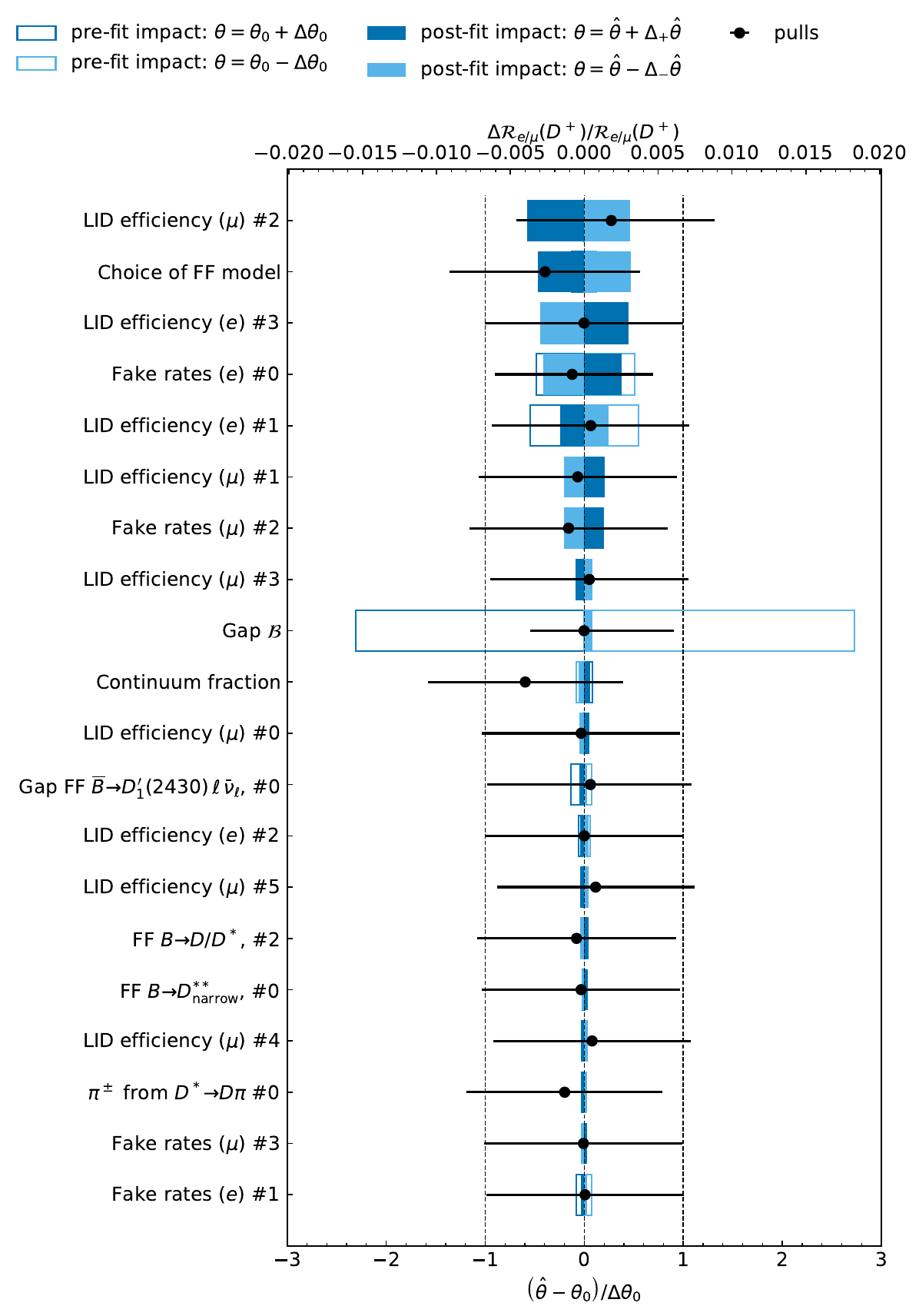}
     \includegraphics[width=0.49\textwidth]{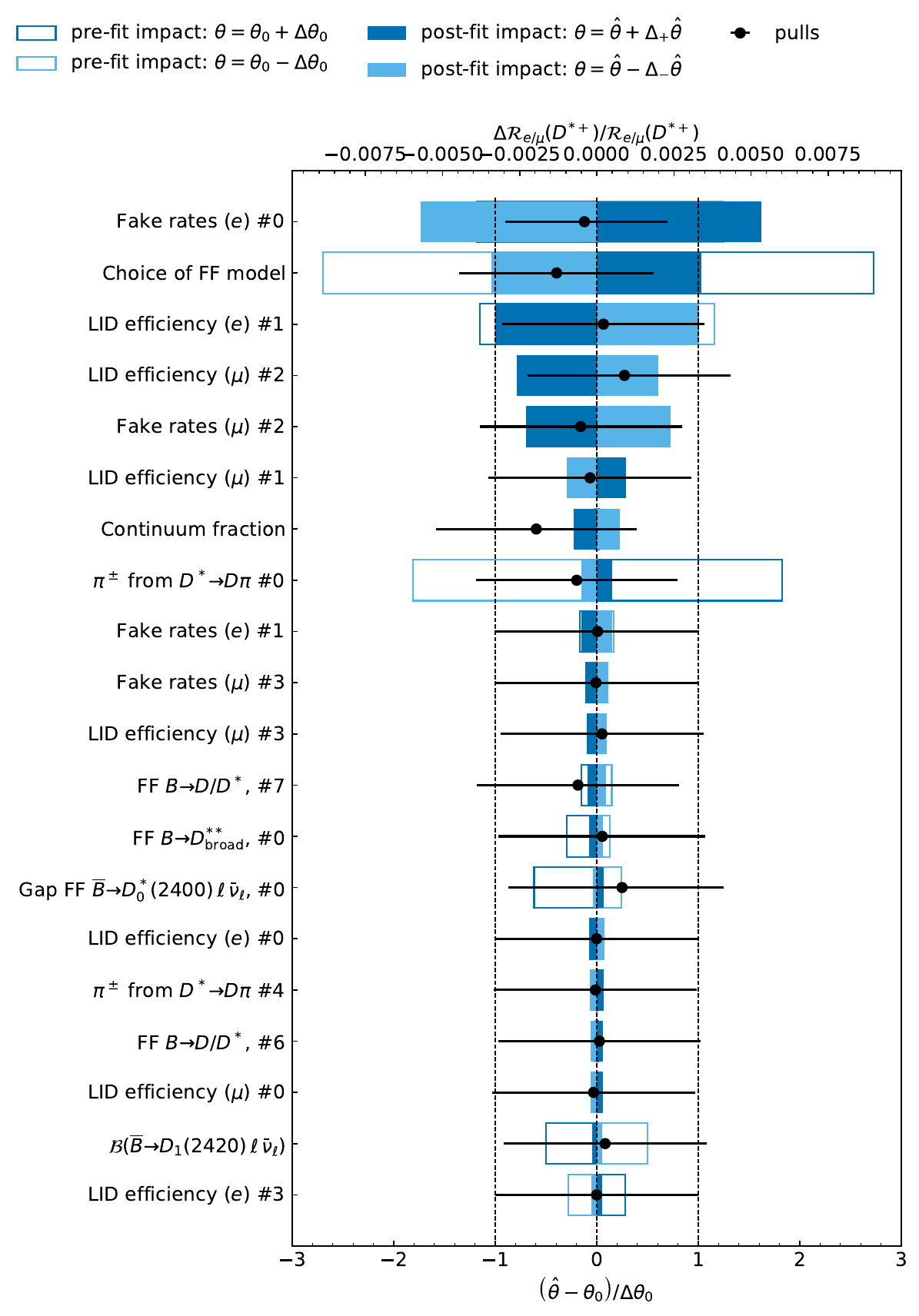}
\caption{Ranking of nuisance parameters for $\mathcal{R}_{e/\mu}(D^+)$ (left) and $\mathcal{R}_{e/\mu}(D^{*+})$ (right). The 20 most impactful nuisance parameters are shown, ordered by their influence on the corresponding fitted ratio. The visual representation follows that of Fig.~\ref{fig:pullsRD}.}
	\label{fig:pullsRDemu}
\end{figure}

Table~\ref{tab:systematics_lfu} presents a breakdown of the systematic uncertainties for the ratios $\mathcal{R}(D^+_{e/\mu})$ and $\mathcal{R}(D^{*+}_{e/\mu})$ into the same sources discussed in Sec.~\ref{sec:systematics}. Fig.~\ref{fig:pullsRDemu} illustrates the 20 most influential nuisance parameters, in a manner similar to Appendix~\ref{app:B}.

\begin{table}[h!]
	\footnotesize
	\renewcommand{\arraystretch}{1.3}
	\begin{center}
		\caption{Systematic uncertainties on $\mathcal{R}(D^+_{e/\mu})$ and $\mathcal{R}(D^{*+}_{e/\mu})$, ranked by their impact on $\mathcal{R}(D^+_{e/\mu})$, following the format of Table~\ref{tab:systematics}}
		\begin{tabular}{lll}
			\hline\hline
			Systematic Uncertainty & $\Delta \mathcal{R}(D^+_{e/\mu})$ &  $\Delta \mathcal{R}(D^{*+}_{e/\mu})$ \\
			\hline
			{\bf Additive} & & \\
			$\,$ MC sample size  & $^{+0.017~(1.5\%)}_{-0.016~(1.6\%)}$   & $^{+0.014~(1.3\%)}_{-0.012~(1.2\%)}$ \\
			$\,$ BDT modeling         & 0.009 (0.8\%)   & 0.007 (0.6\%) \\
		
			$\,$ $\overline{B}{} \to D^{(*)} \ell \bar \nu_\ell$ / $\tau \bar \nu_\tau$  FFs & $0.003~(0.3\%)$   & 0.003 (0.3\%) \\

			$$\,$$ Fake rates ($e$)     & $0.003(0.3\%)$   & $^{+0.005~(0.4\%)}_{-0.006~(0.5\%)}$  \\
			$$\,$$ LID efficiency ($\mu$)  & 0.001 (0.1\%) &  0.001 (0.1\%)  \\
			$\,$ $\overline{B}{} \to D^{**} \ell \bar \nu_\ell$ FFs & 0.001 (0.1\%)   & $0.001~(0.1\%)$ \\
			$$\,$$ Fake rates ($\mu$)     & 0.001 (0.1\%)   & $0.002 ~(0.2\%)$  \\
			$\,$ Gap $\cal B$         & 0.001 (0.1\%) & $ 0.001 ~(0.1\%) $ \\
		
			$\,$ Gap FFs         &  0.001 (0.1\%)  &  $0.001 ~(0.1\%)$ \\
			$\,$ Continuum fraction   &   0.001 (0.1\%)   & 0.001 (0.1\%) \\
			$\,$ $\pi^\pm$ from $D^* \to D \pi$ &  0.001 (0.1\%)   & 0.001 (0.1\%)\\

			$\,$ $\mathcal{B}(\overline{B}{} \to D^{**} \ell \bar \nu_\ell)$ &  0.001 (0.1\%)   & 0.001 (0.1\%) \\
            $$\,$$ LID efficiency ($e$)  & 0.001 (0.1\%)   & 0.003 (0.3\%) \\
			\hline
			{\bf Total Additive Uncertainty} & $^{+0.020~(1.8\%)}_{-0.019 ~(1.7\%)}$  & $^{+0.017 ~(1.6\%)}_{-0.016 ~(1.5\%)}$ \\
			\hline

			{\bf Multiplicative} & & \\
		
			$\,$ MC sample size  & 0.012 (1.1\%)   & 0.009 (0.9\%) \\
			$$\,$$ LID efficiency ($e$)  & 0.004 (0.4\%)   & 0.004 (0.4\%) \\
			
			$$\,$$ LID efficiency ($\mu$)  & 0.003 (0.3\%)   & 0.003 (0.3\%) \\
			
			$\,$ $\overline{B}{} \to D^{(*)} \ell \bar \nu_\ell$ / $\tau \bar \nu_\tau$ FFs & 0.001 (0.1\%)   & 0.001 (0.1\%) \\
			$\, $ $\pi^\pm$ from $D^* \to D \pi$ & -- $\quad\,$ (--)  & 0.001 (0.01\%) \\
			$\,$ Tracking efficiency & 0.001 (0.1\%)  & 0.001 (0.1\%) \\
		
			\hline
			{\bf Total Multiplicative Uncertainty} & 0.013 (1.2\%) & 0.011 (1.2\%) \\
			\hline
			{\bf Total Syst. Uncertainty} & $^{+0.024 (2.2\%)}_{-0.023 (2.2\%)}$  & $^{+0.020 (1.9\%)}_{-0.019 (1.8\%)}$ \\
			\hline \hline
			{\bf Total Stat. Uncertainty} & $^{+ 0.050  (4.7\%)}_{-0.048  (4.5\%)}$  & $^{+ 0.037 (3.4\%) }_{ -0.036 (3.3\%) }$ \\
			\hline
			{\bf Total Uncertainty} &  $^{+0.056 (5.2\%)}_{-0.053 (5.0\%)}$   &$^{+0.042~(3.9\%)}_{-0.041 ~(3.8\%)}$ \\
			\hline\hline
		\end{tabular}
		\label{tab:systematics_lfu}
	\end{center}
\end{table}

\FloatBarrier

%%%%%%%%%%%%%%%%%%%%%%%%%%%%%%%%%%%%%%%%%%%%%%%%%%%%%%%%%%%%%%%%%%%%%%%%%%%%%%%%

\end{appendix}
\end{document}

%% file: pub061.tex
%%% Paper:    LFV with R(D(*))
%%% Journal:  PhysRev
%%% Contacts: A. Manthei
%%% ====================================================================
%%% Use \input{pub061} to insert this material into your latex file.
\newcommand{\instCPPM}{Aix Marseille Universit\'{e}, CNRS/IN2P3, CPPM, 13288 Marseille, France}
\newcommand{\instYerevan}{Alikhanyan National Science Laboratory, Yerevan 0036, Armenia}
\newcommand{\instGoettingen}{II. Physikalisches Institut, Georg-August-Universit\"{a}t G\"{o}ttingen, 37073 G\"{o}ttingen, Germany}
\newcommand{\instBeihang}{Beihang University, Beijing 100191, China}
\newcommand{\instBNL}{Brookhaven National Laboratory, Upton, New York 11973, U.S.A.}
\newcommand{\instBINP}{Budker Institute of Nuclear Physics SB RAS, Novosibirsk 630090, Russian Federation}
\newcommand{\instCMU}{Carnegie Mellon University, Pittsburgh, Pennsylvania 15213, U.S.A.}
\newcommand{\instCinvestavIPN}{Centro de Investigacion y de Estudios Avanzados del Instituto Politecnico Nacional, Mexico City 07360, Mexico}
\newcommand{\instPrague}{Faculty of Mathematics and Physics, Charles University, 121 16 Prague, Czech Republic}
\newcommand{\instChiangMai}{Chiang Mai University, Chiang Mai 50202, Thailand}
\newcommand{\instChiba}{Chiba University, Chiba 263-8522, Japan}
\newcommand{\instChonnam}{Chonnam National University, Gwangju 61186, South Korea}
\newcommand{\instChula}{Chulalongkorn University, Bangkok 10330, Thailand}
\newcommand{\instCAU}{Chung-Ang University, Seoul 06974, South Korea}
\newcommand{\instConacyt}{Consejo Nacional de Ciencia y Tecnolog\'{\i}a, Mexico City 03940, Mexico}
\newcommand{\instDESY}{Deutsches Elektronen--Synchrotron, 22607 Hamburg, Germany}
\newcommand{\instDuke}{Duke University, Durham, North Carolina 27708, U.S.A.}
\newcommand{\instITAR}{Institute of Theoretical and Applied Research (ITAR), Duy Tan University, Hanoi 100000, Vietnam}
\newcommand{\instFuJen}{Department of Physics, Fu Jen Catholic University, Taipei 24205, Taiwan}
\newcommand{\instFudan}{Key Laboratory of Nuclear Physics and Ion-beam Application (MOE) and Institute of Modern Physics, Fudan University, Shanghai 200443, China}
\newcommand{\instGifu}{Gifu University, Gifu 501-1193, Japan}
\newcommand{\instSOKENDAI}{The Graduate University for Advanced Studies (Sokendai), Tsukuba 305-0801, Japan}
\newcommand{\instMainz}{Institut f\"{u}r Kernphysik, Johannes Gutenberg-Universit\"{a}t Mainz, D-55099 Mainz, Germany}
\newcommand{\instGyeongsang}{Gyeongsang National University, Jinju 52828, South Korea}
\newcommand{\instHanyang}{Department of Physics and Institute of Natural Sciences, Hanyang University, Seoul 04763, South Korea}
\newcommand{\instHeidelberg}{Heidelberg University, 68131 Mannheim, Germany}
\newcommand{\instHNU}{Henan Normal University, Xinxiang 453007, China}
\newcommand{\instKEK}{High Energy Accelerator Research Organization (KEK), Tsukuba 305-0801, Japan}
\newcommand{\instJPARC}{J-PARC Branch, KEK Theory Center, High Energy Accelerator Research Organization (KEK), Tsukuba 305-0801, Japan}
\newcommand{\instHiroshima}{Hiroshima University, Higashi-Hiroshima, Hiroshima 739-8530, Japan}
\newcommand{\instHUNNU}{Hunan Normal University, Changsha 410081, China}
\newcommand{\instFrascati}{INFN Laboratori Nazionali di Frascati, I-00044 Frascati, Italy}
\newcommand{\instNapoliINFN}{INFN Sezione di Napoli, I-80126 Napoli, Italy}
\newcommand{\instPadovaINFN}{INFN Sezione di Padova, I-35131 Padova, Italy}
\newcommand{\instPerugiaINFN}{INFN Sezione di Perugia, I-06123 Perugia, Italy}
\newcommand{\instPisaINFN}{INFN Sezione di Pisa, I-56127 Pisa, Italy}
\newcommand{\instRomaTreINFN}{INFN Sezione di Roma Tre, I-00146 Roma, Italy}
\newcommand{\instTorinoINFN}{INFN Sezione di Torino, I-10125 Torino, Italy}
\newcommand{\instTriesteINFN}{INFN Sezione di Trieste, I-34127 Trieste, Italy}
\newcommand{\instIISc}{Indian Institute of Science, Bengaluru 560012, India}
\newcommand{\instIISER}{Indian Institute of Science Education and Research Mohali, SAS Nagar, 140306, India}
\newcommand{\instIITBhubaneswar}{Indian Institute of Technology Bhubaneswar, Bhubaneswar 752050, India}
\newcommand{\instIITGuwahati}{Indian Institute of Technology Guwahati, Assam 781039, India}
\newcommand{\instIITHyderabad}{Indian Institute of Technology Hyderabad, Telangana 502285, India}
\newcommand{\instIITJodhpur}{Indian Institute of Technology Jodhpur, Jodhpur 342030, India}
\newcommand{\instIITMadras}{Indian Institute of Technology Madras, Chennai 600036, India}
\newcommand{\instIndiana}{Indiana University, Bloomington, Indiana 47408, U.S.A.}
\newcommand{\instIHEPRussia}{Institute for High Energy Physics, Protvino 142281, Russian Federation}
\newcommand{\instHEPHYVienna}{Institute of High Energy Physics, Vienna 1050, Austria}
\newcommand{\instIHEPChina}{Institute of High Energy Physics, Chinese Academy of Sciences, Beijing 100049, China}
\newcommand{\instIPP}{Institute of Particle Physics (Canada), Victoria, British Columbia V8W 2Y2, Canada}
\newcommand{\instIOP}{Institute of Physics, Vietnam Academy of Science and Technology (VAST), Hanoi, Vietnam}
\newcommand{\instIFIC}{Instituto de Fisica Corpuscular, Paterna 46980, Spain}
\newcommand{\instISU}{Iowa State University, Ames, Iowa 50011, U.S.A.}
\newcommand{\instJAEA}{Advanced Science Research Center, Japan Atomic Energy Agency, Naka 319-1195, Japan}
\newcommand{\instJLU}{Jilin University, Changchun 130012, China}
\newcommand{\instKarlsruhe}{Institut f\"{u}r Experimentelle Teilchenphysik, Karlsruher Institut f\"{u}r Technologie, 76131 Karlsruhe, Germany}
\newcommand{\instKitasato}{Kitasato University, Sagamihara 252-0373, Japan}
\newcommand{\instKISTI}{Korea Institute of Science and Technology Information, Daejeon 34141, South Korea}
\newcommand{\instKoreaUnivKU}{Korea University, Seoul 02841, South Korea}
\newcommand{\instKSU}{Kyoto Sangyo University, Kyoto 603-8555, Japan}
\newcommand{\instKyotoU}{Kyoto University, Kyoto 606-8501, Japan}
\newcommand{\instKyungpook}{Kyungpook National University, Daegu 41566, South Korea}
\newcommand{\instLNNU}{Liaoning Normal University, Dalian 116029, China}
\newcommand{\instGiessen}{Justus-Liebig-Universit\"{a}t Gie\ss{}en, 35392 Gie\ss{}en, Germany}
\newcommand{\instLuther}{Luther College, Decorah, Iowa 52101, U.S.A.}
\newcommand{\instMNITJaipur}{Malaviya National Institute of Technology Jaipur, Jaipur 302017, India}
\newcommand{\instLMU}{Ludwig Maximilians University, 80539 Munich, Germany}
\newcommand{\instMcGill}{McGill University, Montr\'{e}al, Qu\'{e}bec, H3A 2T8, Canada}
\newcommand{\instMETU}{Middle East Technical University, 06531 Ankara, Turkey}
\newcommand{\instMEPhI}{Moscow Physical Engineering Institute, Moscow 115409, Russian Federation}
\newcommand{\instNagoya}{Graduate School of Science, Nagoya University, Nagoya 464-8602, Japan}
\newcommand{\instNagoyaKMI}{Kobayashi-Maskawa Institute, Nagoya University, Nagoya 464-8602, Japan}
\newcommand{\instNankai}{Nankai University, Tianjin 300071, China}
\newcommand{\instNaraWu}{Nara Women's University, Nara 630-8506, Japan}
\newcommand{\instUNAM}{National Autonomous University of Mexico, Mexico City, Mexico}
\newcommand{\instHSE}{National Research University Higher School of Economics, Moscow 101000, Russian Federation}
\newcommand{\instNTUTaiwan}{Department of Physics, National Taiwan University, Taipei 10617, Taiwan}
\newcommand{\instNUUTaiwan}{National United University, Miao Li 36003, Taiwan}
\newcommand{\instKrakow}{H. Niewodniczanski Institute of Nuclear Physics, Krakow 31-342, Poland}
\newcommand{\instNiigata}{Niigata University, Niigata 950-2181, Japan}
\newcommand{\instNSU}{Novosibirsk State University, Novosibirsk 630090, Russian Federation}
\newcommand{\instOsakaMetropolitan}{Osaka Metropolitan University, Osaka 558-8585, Japan}
\newcommand{\instRCNP}{Research Center for Nuclear Physics, Osaka University, Osaka 567-0047, Japan}
\newcommand{\instPNNL}{Pacific Northwest National Laboratory, Richland, Washington 99352, U.S.A.}
\newcommand{\instPanjab}{Panjab University, Chandigarh 160014, India}
\newcommand{\instMPP}{Max-Planck-Institut f\"{u}r Physik, 80805 M\"{u}nchen, Germany}
\newcommand{\instMPGHLL}{Semiconductor Laboratory of the Max Planck Society, 81739 M\"{u}nchen, Germany}
\newcommand{\instPanjabPAU}{Punjab Agricultural University, Ludhiana 141004, India}
\newcommand{\instRIKENMSL}{Meson Science Laboratory, Cluster for Pioneering Research, RIKEN, Saitama 351-0198, Japan}
\newcommand{\instSeoul}{Seoul National University, Seoul 08826, South Korea}
\newcommand{\instShandong}{Shandong University, Jinan 250100, China}
\newcommand{\instKyiv}{Taras Shevchenko National University of Kiev, Kiev, Ukraine}
\newcommand{\instSPU}{Showa Pharmaceutical University, Tokyo 194-8543, Japan}
\newcommand{\instSoochow}{Soochow University, Suzhou 215006, China}
\newcommand{\instSoongsil}{Department of Physics and Origin of Matter and Evolution of Galaxy (OMEG) Institute, Soongsil University, Seoul 06978, South Korea}
\newcommand{\instSEU}{Southeast University, Nanjing 211189, China}
\newcommand{\instLjubljanaJSI}{J. Stefan Institute, 1000 Ljubljana, Slovenia}
\newcommand{\instSKKU}{Sungkyunkwan University, Suwon 16419, South Korea}
\newcommand{\instTata}{Tata Institute of Fundamental Research, Mumbai 400005, India}
\newcommand{\instTUM}{Department of Physics, Technische Universit\"{a}t M\"{u}nchen, 85748 Garching, Germany}
\newcommand{\instTelAviv}{School of Physics and Astronomy, Tel Aviv University, Tel Aviv 69978, Israel}
\newcommand{\instToho}{Toho University, Funabashi 274-8510, Japan}
\newcommand{\instTitech}{Tokyo Institute of Technology, Tokyo 152-8550, Japan}
\newcommand{\instTokyoMetropolitan}{Tokyo Metropolitan University, Tokyo 192-0397, Japan}
\newcommand{\instUAS}{Universidad Autonoma de Sinaloa, Sinaloa 80000, Mexico}
\newcommand{\instNapoliUNIV}{Dipartimento di Scienze Fisiche, Universit\`{a} di Napoli Federico II, I-80126 Napoli, Italy}
\newcommand{\instPadovaUNIV}{Dipartimento di Fisica e Astronomia, Universit\`{a} di Padova, I-35131 Padova, Italy}
\newcommand{\instPerugiaUNIV}{Dipartimento di Fisica, Universit\`{a} di Perugia, I-06123 Perugia, Italy}
\newcommand{\instPisaUNIV}{Dipartimento di Fisica, Universit\`{a} di Pisa, I-56127 Pisa, Italy}
\newcommand{\instRomaTreUNIV}{Dipartimento di Matematica e Fisica, Universit\`{a} di Roma Tre, I-00146 Roma, Italy}
\newcommand{\instTorinoUNIV}{Dipartimento di Fisica, Universit\`{a} di Torino, I-10125 Torino, Italy}
\newcommand{\instTriesteUNIV}{Dipartimento di Fisica, Universit\`{a} di Trieste, I-34127 Trieste, Italy}
\newcommand{\instIJCLab}{Universit\'{e} Paris-Saclay, CNRS/IN2P3, IJCLab, 91405 Orsay, France}
\newcommand{\instIPHC}{Universit\'{e} de Strasbourg, CNRS, IPHC, UMR 7178, 67037 Strasbourg, France}
\newcommand{\instAdelaide}{Department of Physics, University of Adelaide, Adelaide, South Australia 5005, Australia}
\newcommand{\instUofA}{University of Alberta, Edmonton, Alberta, T6G 2E1, Canada}
\newcommand{\instBonn}{University of Bonn, 53115 Bonn, Germany}
\newcommand{\instUBC}{University of British Columbia, Vancouver, British Columbia, V6T 1Z1, Canada}
\newcommand{\instCincinnati}{University of Cincinnati, Cincinnati, Ohio 45221, U.S.A.}
\newcommand{\instFlorida}{University of Florida, Gainesville, Florida 32611, U.S.A.}
\newcommand{\instHamburg}{University of Hamburg, 20148 Hamburg, Germany}
\newcommand{\instHawaii}{University of Hawaii, Honolulu, Hawaii 96822, U.S.A.}
\newcommand{\instLjubljanaUniLJ}{Faculty of Mathematics and Physics, University of Ljubljana, 1000 Ljubljana, Slovenia}
\newcommand{\instLouisville}{University of Louisville, Louisville, Kentucky 40292, U.S.A.}
\newcommand{\instMalaya}{National Centre for Particle Physics, University of Malaya, 50603 Kuala Lumpur, Malaysia}
\newcommand{\instManitoba}{University of Manitoba, Winnipeg, Manitoba, R3T 2N2, Canada}
\newcommand{\instLjubljanaUM}{Faculty of Chemistry and Chemical Engineering, University of Maribor, 2000 Maribor, Slovenia}
\newcommand{\instMelbourne}{School of Physics, University of Melbourne, Victoria 3010, Australia}
\newcommand{\instMississippi}{University of Mississippi, University, Mississippi 38677, U.S.A.}
\newcommand{\instUOM}{University of Miyazaki, Miyazaki 889-2192, Japan}
\newcommand{\instNovaGorica}{University of Nova Gorica, 5000 Nova Gorica, Slovenia}
\newcommand{\instUPES}{University of Petroleum and Energy Studies, Dehradun 248007, India}
\newcommand{\instPittsburgh}{University of Pittsburgh, Pittsburgh, Pennsylvania 15260, U.S.A.}
\newcommand{\instUSTC}{Department of Modern Physics and State Key Laboratory of Particle Detection and Electronics, University of Science and Technology of China, Hefei 230026, China}
\newcommand{\instSAlabama}{University of South Alabama, Mobile, Alabama 36688, U.S.A.}
\newcommand{\instSCarolina}{University of South Carolina, Columbia, South Carolina 29208, U.S.A.}
\newcommand{\instSydney}{School of Physics, University of Sydney, New South Wales 2006, Australia}
\newcommand{\instTabuk}{Department of Physics, Faculty of Science, University of Tabuk, Tabuk 71451, Saudi Arabia}
\newcommand{\instUTokyo}{Department of Physics, University of Tokyo, Tokyo 113-0033, Japan}
\newcommand{\instEri}{Earthquake Research Institute, University of Tokyo, Tokyo 113-0032, Japan}
\newcommand{\instIPMU}{Kavli Institute for the Physics and Mathematics of the Universe (WPI), University of Tokyo, Kashiwa 277-8583, Japan}
\newcommand{\instVictoria}{University of Victoria, Victoria, British Columbia, V8W 3P6, Canada}
\newcommand{\instUppsala}{Uppsala University, SE-751 05 Uppsala, Sweden}
\newcommand{\instVPI}{Virginia Polytechnic Institute and State University, Blacksburg, Virginia 24061, U.S.A.}
\newcommand{\instWayneState}{Wayne State University, Detroit, Michigan 48202, U.S.A.}
\newcommand{\instXavier}{St. Francis Xavier University, Antigonish, Nova Scotia, B2G 2W5, Canada}
\newcommand{\instXJTU}{Xi'an Jiaotong University, Xi'an 710049, China}
\newcommand{\instYamagata}{Yamagata University, Yamagata 990-8560, Japan}
\newcommand{\instYonsei}{Yonsei University, Seoul 03722, South Korea}
\newcommand{\instTYL}{Toshiko Yuasa Laboratory, Tsukuba 305-0801, Japan}
\newcommand{\instZZU}{Zhengzhou University, Zhengzhou 450001, China}
\affiliation{\instCPPM}
\affiliation{\instYerevan}
\affiliation{\instGoettingen}
\affiliation{\instBeihang}
\affiliation{\instBNL}
\affiliation{\instBINP}
\affiliation{\instCMU}
\affiliation{\instCinvestavIPN}
\affiliation{\instPrague}
\affiliation{\instChiangMai}
%%%\affiliation{\instChiba}
\affiliation{\instChonnam}
\affiliation{\instChula}
\affiliation{\instCAU}
\affiliation{\instConacyt}
\affiliation{\instDESY}
\affiliation{\instDuke}
\affiliation{\instITAR}
\affiliation{\instFuJen}
\affiliation{\instFudan}
\affiliation{\instGifu}
\affiliation{\instSOKENDAI}
\affiliation{\instMainz}
\affiliation{\instGyeongsang}
\affiliation{\instHanyang}
%%%\affiliation{\instHeidelberg}
\affiliation{\instHNU}
\affiliation{\instKEK}
\affiliation{\instJPARC}
%%%\affiliation{\instHiroshima}
\affiliation{\instHUNNU}
\affiliation{\instFrascati}
\affiliation{\instNapoliINFN}
\affiliation{\instPadovaINFN}
\affiliation{\instPerugiaINFN}
\affiliation{\instPisaINFN}
\affiliation{\instRomaTreINFN}
\affiliation{\instTorinoINFN}
\affiliation{\instTriesteINFN}
\affiliation{\instIISc}
\affiliation{\instIISER}
\affiliation{\instIITBhubaneswar}
\affiliation{\instIITGuwahati}
\affiliation{\instIITHyderabad}
\affiliation{\instIITJodhpur}
\affiliation{\instIITMadras}
\affiliation{\instIndiana}
\affiliation{\instIHEPRussia}
\affiliation{\instHEPHYVienna}
\affiliation{\instIHEPChina}
\affiliation{\instIPP}
\affiliation{\instIOP}
\affiliation{\instIFIC}
\affiliation{\instISU}
\affiliation{\instJAEA}
\affiliation{\instJLU}
\affiliation{\instKarlsruhe}
\affiliation{\instKitasato}
\affiliation{\instKISTI}
\affiliation{\instKoreaUnivKU}
%%%\affiliation{\instKSU}
\affiliation{\instKyotoU}
\affiliation{\instKyungpook}
\affiliation{\instLNNU}
\affiliation{\instGiessen}
\affiliation{\instLuther}
\affiliation{\instMNITJaipur}
\affiliation{\instLMU}
\affiliation{\instMcGill}
\affiliation{\instMETU}
%%%\affiliation{\instMEPhI}
\affiliation{\instNagoya}
\affiliation{\instNagoyaKMI}
\affiliation{\instNankai}
\affiliation{\instNaraWu}
%%%\affiliation{\instUNAM}
\affiliation{\instHSE}
\affiliation{\instNTUTaiwan}
%%%\affiliation{\instNUUTaiwan}
\affiliation{\instKrakow}
\affiliation{\instNiigata}
\affiliation{\instNSU}
\affiliation{\instOsakaMetropolitan}
\affiliation{\instRCNP}
\affiliation{\instPNNL}
\affiliation{\instPanjab}
\affiliation{\instMPP}
%%%\affiliation{\instMPGHLL}
%%%\affiliation{\instPanjabPAU}
\affiliation{\instRIKENMSL}
%%%\affiliation{\instSeoul}
\affiliation{\instShandong}
\affiliation{\instKyiv}
\affiliation{\instSPU}
\affiliation{\instSoochow}
\affiliation{\instSoongsil}
\affiliation{\instSEU}
\affiliation{\instLjubljanaJSI}
\affiliation{\instSKKU}
\affiliation{\instTata}
\affiliation{\instTUM}
\affiliation{\instTelAviv}
\affiliation{\instToho}
\affiliation{\instTitech}
%%%\affiliation{\instTokyoMetropolitan}
\affiliation{\instUAS}
\affiliation{\instNapoliUNIV}
\affiliation{\instPadovaUNIV}
\affiliation{\instPerugiaUNIV}
\affiliation{\instPisaUNIV}
\affiliation{\instRomaTreUNIV}
\affiliation{\instTorinoUNIV}
\affiliation{\instTriesteUNIV}
\affiliation{\instIJCLab}
\affiliation{\instIPHC}
\affiliation{\instAdelaide}
\affiliation{\instUofA}
\affiliation{\instBonn}
\affiliation{\instUBC}
\affiliation{\instCincinnati}
\affiliation{\instFlorida}
\affiliation{\instHamburg}
\affiliation{\instHawaii}
\affiliation{\instLjubljanaUniLJ}
\affiliation{\instLouisville}
%%%\affiliation{\instMalaya}
\affiliation{\instManitoba}
\affiliation{\instLjubljanaUM}
\affiliation{\instMelbourne}
\affiliation{\instMississippi}
%%%\affiliation{\instUOM}
%%%\affiliation{\instNovaGorica}
\affiliation{\instUPES}
\affiliation{\instPittsburgh}
\affiliation{\instUSTC}
\affiliation{\instSAlabama}
%%%\affiliation{\instSCarolina}
\affiliation{\instSydney}
\affiliation{\instTabuk}
\affiliation{\instUTokyo}
\affiliation{\instEri}
\affiliation{\instIPMU}
\affiliation{\instVictoria}
\affiliation{\instUppsala}
\affiliation{\instVPI}
\affiliation{\instWayneState}
\affiliation{\instXavier}
\affiliation{\instXJTU}
\affiliation{\instYamagata}
\affiliation{\instYonsei}
\affiliation{\instTYL}
\affiliation{\instZZU}
  \author{I.~Adachi}\affiliation{\instKEK}\affiliation{\instSOKENDAI} % 2590
  \author{K.~Adamczyk}\affiliation{\instLjubljanaUniLJ}\affiliation{\instLjubljanaJSI} % 2239
  \author{L.~Aggarwal}\affiliation{\instPanjab} % 10083
% \author{P.~Ahlburg}\affiliation{\instBonn} % 2367
  \author{H.~Ahmed}\affiliation{\instXavier} % 11323
% \author{J.~K.~Ahn}\affiliation{\instKoreaUnivKU} % 7423
% \author{Y.~Ahn}\affiliation{\instKoreaUnivKU} % 14363
  \author{H.~Aihara}\affiliation{\instUTokyo} % 2223
  \author{N.~Akopov}\affiliation{\instYerevan} % 9443
  \author{S.~Alghamdi}\affiliation{\instTabuk} % 27804
  \author{M.~Alhakami}\affiliation{\instTabuk} % 28103
  \author{A.~Aloisio}\affiliation{\instNapoliUNIV}\affiliation{\instNapoliINFN} % 2194
  \author{N.~Althubiti}\affiliation{\instTabuk} % 21524
% ...
% 13403
% \author{S.~Zou}\affiliation{\instFudan} % 19363
\collaboration{The Belle II Collaboration}

%% file: acknowledgements-b2.tex
% Policy from October 20, 2022
This work, based on data collected using the Belle II detector, which was built and commissioned prior to March 2019,
%Belle1 and data collected using the Belle detector, which was operated until June 2010,
was supported by
%Armenia
Higher Education and Science Committee of the Republic of Armenia Grant No.~23LCG-1C011;
%Australia
Australian Research Council and Research Grants
No.~DP200101792, % Jackson
No.~DP210101900, % Urquijo
No.~DP210102831, % Sevior
No.~DE220100462, % Hsu
No.~LE210100098, % Infrastructure
and
No.~LE230100085; % Infrastructure
%Austria
Austrian Federal Ministry of Education, Science and Research,
Austrian Science Fund (FWF) Grants
DOI:~10.55776/P34529,
DOI:~10.55776/J4731,
DOI:~10.55776/J4625,
DOI:~10.55776/M3153,
and
DOI:~10.55776/PAT1836324,
and
Horizon 2020 ERC Starting Grant No.~947006 ``InterLeptons'';
%Canada
Natural Sciences and Engineering Research Council of Canada, Compute Canada and CANARIE;
%China
National Key R\&D Program of China under Contract No.~2024YFA1610503,
and
No.~2024YFA1610504
National Natural Science Foundation of China and Research Grants
No.~11575017,
No.~11761141009,
No.~11705209,
No.~11975076,
No.~12135005,
No.~12150004,
No.~12161141008,
No.~12475093,
and
No.~12175041,
and Shandong Provincial Natural Science Foundation Project~ZR2022JQ02;
%Czech Republic
the Czech Science Foundation Grant No.~22-18469S 
and
Charles University Grant Agency project No.~246122;
%EU
European Research Council, Seventh Framework PIEF-GA-2013-622527,
Horizon 2020 ERC-Advanced Grants No.~267104 and No.~884719,
Horizon 2020 ERC-Consolidator Grant No.~819127,
Horizon 2020 Marie Sklodowska-Curie Grant Agreement No.~700525 ``NIOBE''
and
No.~101026516,
and
Horizon 2020 Marie Sklodowska-Curie RISE project JENNIFER2 Grant Agreement No.~822070 (European grants);
%France
L'Institut National de Physique Nucl\'{e}aire et de Physique des Particules (IN2P3) du CNRS
and
L'Agence Nationale de la Recherche (ANR) under Grant No.~ANR-21-CE31-0009 (France);
%Germany
BMBF, DFG, HGF, MPG, and AvH Foundation (Germany);
%India
Department of Atomic Energy under Project Identification No.~RTI 4002,
Department of Science and Technology,
and
UPES SEED funding programs
No.~UPES/R\&D-SEED-INFRA/17052023/01 and
No.~UPES/R\&D-SOE/20062022/06 (India);
%Israel
Israel Science Foundation Grant No.~2476/17,
U.S.-Israel Binational Science Foundation Grant No.~2016113, and
Israel Ministry of Science Grant No.~3-16543;
%Italy
Istituto Nazionale di Fisica Nucleare and the Research Grants BELLE2,
and
the ICSC – Centro Nazionale di Ricerca in High Performance Computing, Big Data and Quantum Computing, funded by European Union – NextGenerationEU;
%Japan
Japan Society for the Promotion of Science, Grant-in-Aid for Scientific Research Grants
No.~16H03968,
No.~16H03993,
No.~16H06492,
No.~16K05323,
No.~17H01133,
No.~17H05405,
No.~18K03621,
No.~18H03710,
No.~18H05226,
No.~19H00682, % Niigata
No.~20H05850,
No.~20H05858,
No.~22H00144,
No.~22K14056,
No.~22K21347,
No.~23H05433,
No.~26220706,
and
No.~26400255,
%the National Institute of Informatics, and Science Information NETwork 5 (SINET5), 
and
the Ministry of Education, Culture, Sports, Science, and Technology (MEXT) of Japan;  
%Korea
National Research Foundation (NRF) of Korea Grants
No.~2016R1-D1A1B-02012900,
No.~2018R1-A6A1A-06024970,
No.~2021R1-A6A1A-03043957,
No.~2021R1-F1A-1060423,
No.~2021R1-F1A-1064008,
No.~2022R1-A2C-1003993,
No.~2022R1-A2C-1092335,
No.~RS-2023-00208693,
No.~RS-2024-00354342
and
No.~RS-2022-00197659,
Radiation Science Research Institute,
Foreign Large-Size Research Facility Application Supporting project,
the Global Science Experimental Data Hub Center, the Korea Institute of
Science and Technology Information (K24L2M1C4)
and
KREONET/GLORIAD;
%Malaysia
Universiti Malaya RU grant, Akademi Sains Malaysia, and Ministry of Education Malaysia;
%Mexico
% CINVESTAV-IPN, UNAM, UAS, BUAP and CONACYT are funded under
Frontiers of Science Program Contracts
No.~FOINS-296,
No.~CB-221329,
No.~CB-236394,
No.~CB-254409,
and
No.~CB-180023, and SEP-CINVESTAV Research Grant No.~237 (Mexico);
%Poland
the Polish Ministry of Science and Higher Education and the National Science Center;
%Russia
the Ministry of Science and Higher Education of the Russian Federation
and
the HSE University Basic Research Program, Moscow;
%Saudi Arabia
University of Tabuk Research Grants
No.~S-0256-1438 and No.~S-0280-1439 (Saudi Arabia), and
Researchers Supporting Project number (RSPD2025R873), King Saud University, Riyadh,
Saudi Arabia;
%Slovenia
Slovenian Research Agency and Research Grants
No.~J1-9124
and
No.~P1-0135;
%Spain
Ikerbasque, Basque Foundation for Science,
State Agency for Research of the Spanish Ministry of Science and Innovation through Grant No. PID2022-136510NB-C33, Spain,
Agencia Estatal de Investigacion, Spain
Grant No.~RYC2020-029875-I
and
Generalitat Valenciana, Spain
Grant No.~CIDEGENT/2018/020;
%Swiss (Belle 1)
%Belle1 the Swiss National Science Foundation;
%Sweden
The Knut and Alice Wallenberg Foundation (Sweden), Contracts No.~2021.0174 and No.~2021.0299;
%Taiwan
National Science and Technology Council,
and
Ministry of Education (Taiwan);
%Thailand
Thailand Center of Excellence in Physics;
%Turkey
TUBITAK ULAKBIM (Turkey);
%Ukraine
National Research Foundation of Ukraine, Project No.~2020.02/0257,
and
Ministry of Education and Science of Ukraine;
%USA
the U.S. National Science Foundation and Research Grants
No.~PHY-1913789 % Indiana CEEM
and
No.~PHY-2111604, % Luther
and the U.S. Department of Energy and Research Awards
No.~DE-AC06-76RLO1830, % PNNL
No.~DE-SC0007983, % Wayne State
No.~DE-SC0009824, % Florida
No.~DE-SC0009973, % VPI
No.~DE-SC0010007, % Duke
No.~DE-SC0010073, % South Carolina
No.~DE-SC0010118, % Carnegie Mellon
No.~DE-SC0010504, % Hawaii
No.~DE-SC0011784, % Cincinnati
No.~DE-SC0012704, % BNL
No.~DE-SC0019230, % Duke
No.~DE-SC0021274, % Mississippi
No.~DE-SC0021616, % Mississippi
No.~DE-SC0022350, % Louisville
No.~DE-SC0023470; % South Alabama
%last group
and
%Vietnam
the Vietnam Academy of Science and Technology (VAST) under Grants
No.~NVCC.05.12/22-23
and
No.~DL0000.02/24-25.

% Policy from October 20, 2022
These acknowledgements are not to be interpreted as an endorsement of any statement made
by any of our institutes, funding agencies, governments, or their representatives.

We thank the SuperKEKB team for delivering high-luminosity collisions;
the KEK cryogenics group for the efficient operation of the detector solenoid magnet and IBBelle on site;
the KEK Computer Research Center for on-site computing support; the NII for SINET6 network support;
and the raw-data centers hosted by BNL, DESY, GridKa, IN2P3, INFN, 
%Belle1 PNNL/EMSL, 
and the University of Victoria.